\documentclass[preprint]{aastex}
\usepackage{natbib}
\usepackage{hyperref}
\usepackage{fancyhdr}
\usepackage{graphicx}
\usepackage{epstopdf}
\newcommand{\vs}           {{\it vs.}}

\newcommand{\asec}      {\mbox{$^{\prime\prime}$}}

\newcommand{\be}           {\begin{equation}}
\newcommand{\ee}           {\end{equation}}
\newcommand{\bea}          {\begin{eqnarray}}
\newcommand{\eea}          {\end{eqnarray}}
\newcommand{\WISE}       {{\sl WISE}}
\newcommand{\Var} {\mathrm{Var}}

\slugcomment{3 Nov 2018}

\shorttitle{Response to M2018b}
\shortauthors{Wright et al.}

\begin{document}

\title{Response to ``An empirical examination of WISE/NEOWISE asteroid
analysis and results''}

\author{Edward L.\ Wright\altaffilmark{1}}
\affil{UCLA Astronomy, PO Box 951547, Los Angeles CA 90095-1547}
\and
\author{Amy Mainzer\altaffilmark{2}}
\author{Joseph Masiero\altaffilmark{2}}
\author{Tommy Grav\altaffilmark{3}}
\author{Roc M. Cutri\altaffilmark{4}}
\author{James Bauer\altaffilmark{2,4}}

\altaffiltext{1}{wright@astro.ucla.edu}
\altaffiltext{2}{Jet Propulsion Laboratory, California Institute of Technology,
4800 Oak Grove Dr., Pasadena, CA, 91109-8001, USA}
\altaffiltext{3}{Planetary Sciences Institute, 1700 E Fort Lowell Rd \#106, Tucson, AZ 85719}
\altaffiltext{4}{California Institute of Technology, IPAC, 1200 E.\ California Blvd. Pasadena, CA 92115}

\begin{abstract}

In this paper, we show that a number of claims made in \citet{myhrvold:2018b} (hereafter M2018b) regarding the WISE data and thermal modeling of asteroids are incorrect. That paper provides thermal fit parameter outputs for only two of the $\sim$150,000 object dataset and does not make a direct comparison to asteroids with diameters measured by other means to assess the quality of that work's thermal model. We are unable to reproduce the results for the two objects for which M2018b published its own thermal fit outputs, including diameter, albedo, beaming, and infrared albedo. In particular, the infrared albedos published in M2018b are unphysically low.  Except for these two objects, M2018b does not  publish its own table of thermal fit parameters, nor does it compare these fitted results to diameters measured by other techniques such as occultation or radar, or other missions such as IRAS and AKARI.

While there were some minor issues with consistency between tables due to clerical errors in the WISE/NEOWISE team's various papers and data release in the Planetary Data System, and a software issue that slightly increased diameter uncertainties in some cases, these issues do not substantially change the results and conclusions drawn from the data. We have shown in previous work and with updated analyses presented here that the effective spherical diameters for asteroids published to date are accurate to within the previously quoted minimum systematic 1-sigma uncertainty of $\sim$10\% when data of appropriate quality and quantity are available. Moreover, we show that the method used by M2018b to compare diameters between various asteroid  datasets is incorrect and overestimates their differences. In addition, among other misconceptions in M2018b, we show that the WISE photometric measurement uncertainties are appropriately characterized and used by the WISE data processing pipeline and NEOWISE thermal modeling software. We show that the Near-Earth Asteroid Thermal Model \citep{harris:1998} employed by the NEOWISE team is a very useful model for analyzing infrared data to derive diameters and albedos when used properly.

\end{abstract}

\section{Introduction}

The Wide-field Infrared Survey Explorer [WISE]  \citep{wright/etal:2010} was launched on 14 Dec 2009,
and surveyed the entire sky in 4 infrared bands at 3.4, 4.6, 12 and 22 $\mu$m (denoted W1, W2, W3, and W4 respectively).
The initial orbital period was 95.2 minutes, and the survey strategy was to scan
at approximately the orbital rate in order to always point away from the Earth.  The motion of the spacecraft was compensated for by scanning a small mirror in the optical system at a particular rate to keep the images from streaking. The exact scan rate of the spacecraft was chosen to match the mirror's scan rate in order to ``freeze'' the sky image on the detectors for 9.9 sec out of the
11 sec period between exposures.  
The final scan rate of 3.7966 arc-min/sec produces a 41.8 arc-min
spacing between frames.  Since the field-of-view of the camera is 47 arc-min,
this spacing lead to a 5 arc-min overlap between consecutive frames.

WISE was originally designed to conduct an all-sky survey in all four channels in six months. The observatory met this primary mission objective and continued to survey in all 4 bands
until 6 Aug 2010, when the solid hydrogen cryogen in the large tank that cooled the telescope was exhausted.   This 4 band portion of the WISE
mission (the so-called ``fully cryogenic" phase) makes up
the All-Sky data release which provides the individual exposures as calibrated
FITS files with astrometric solutions, along with a database of sources
detected in these exposures\footnote{http://irsa.ipac.caltech.edu/Missions/wise.html}.
The optics then warmed to 45 K, which
led to a telescope thermal emission that saturated the 22 $\mu$m band.
The hydrogen in the small inner tank that cooled the 12 and 22 $\mu$m detectors lasted until
29 Sep 2010, which allowed the 12 $\mu$m detector to continue operating, even though
it too suffered from greatly increased background which required reducing the
exposure time in steps  to 4.4, then 2.2, and finally 1.1 sec instead of the original 8.8 sec.
With both cryogen tanks empty, the detectors then warmed to 74 K, which still
allowed the 3.4 \& 4.6  $\mu$m detectors to operate.  WISE operated in this
2-band mode from October 2010 through January 2011 (called the post-cryogenic phase), and then was placed into
hibernation.  WISE was reactivated in late 2013 as the NEOWISE reactivation mission using the remaining W1 and W2 channels \citep{mainzer/etal:2014}, and started surveying on 13 Dec 2013; the survey 
continues into 2018, although
the decay of the orbit is leading to gradually increasing focal plane temperatures,
close to 77 K in the summer of 2018. The  individual exposures as calibrated FITS files with astrometric solutions, along with a database of sources detected in these exposures, have been released for all these phases of the mission
through 13 Dec 2017  \citep{cutri/etal:2012,
cutri/etal:2015}, with a final release to be scheduled after the end of survey operations\footnote{Data from all phases of WISE/NEOWISE are available at http://irsa.ipac.caltech.edu/Missions/wise.html.}.

The NEOWISE team has analyzed the infrared observations of asteroids to
generate a very large  dataset of diameters and other physical properties for of order $\sim$150,000 objects
\citep{mainzer/etal:2016} using the Near Earth Asteroid Thermal
Model (NEATM) described by \citet{harris:1998}.  The NEATM is a very simple model that is
easy to apply to large datasets, unlike computationally intensive thermophysical models
\citep[e.g.][]{hanus/etal:2018, rozitis:2018}.

\section{The Quality of the WISE/NEOWISE Results}

\subsection{The 2011 Calibration Method}

\citet{mainzer/etal:2011a} used the NEATM to see
whether the surface brightnesses of asteroids and natural satellites with known diameters
could be reproduced in order to verify the calibration of the then-newly-launched WISE mission's bandpasses for extremely red objects. This set of 117 objects was drawn from radar, occultation, and spacecraft (ROS) data.
In the \citet{mainzer/etal:2016} typology for WISE/NEOWISE diameters archived in NASA's Planetary Data System, these are denoted ``-VBI'' or ``-VB-" fits (where V, B, and I denote visible albedo, beaming, and infrared albedo being fit respectively; the leading dash indicates that diameter was not fit but held fixed).  Asteroids have red spectral energy distributions (SEDs), and the WISE bandpasses are broad. \citet{mainzer/etal:2011a} sought to verify that the zeropoints and color corrections derived for the WISE bandpasses from calibrator stars (which are blue) and Active Galactic Nuclei (which tend to be red but can be variable) were appropriate for objects with very red SEDs such as asteroids. To that end, the differences between model and observed magnitudes 
were plotted \vs\ the asteroid calibrator objects' sub-solar temperatures when their diameters were held fixed to previously published values in order to verify that the newly derived color corrections worked properly for these asteroids, which are much cooler than stars. The resulting parameters are shown in Table 1 of \citet{mainzer/etal:2011a}, and the
deviations in surface brightness are plotted in Figure 3 of that paper. 

\citet{mainzer/etal:2011a} found that the deviations between model magnitudes and measured WISE magnitudes
were small. Since an asteroid's flux goes as the square of its diameter, the minimum diameter uncertainty must scale as one-half of the flux uncertainty. But the actual diameter uncertainty implied can be somewhat larger due to
the correlation of diameter $D$ with the NEATM beaming parameter $\eta$, a model parameter used as a ``catch-all" term to account for a variety of unknown properties of an asteroid's surface, rotational state, observational geometry, and shape. Since the point of the analysis in \citet{mainzer/etal:2011a} was to verify the then-newly-derived zero points and color corrections for the four WISE bands, a plot of the previously measured diameters of the calibration objects  \vs\ the diameters derived for the objects using WISE data was not shown for that set of observations. 

However, such a plot was published for the post-cryogenic and NEOWISE reactivation data using channels W1 and W2 only \citep{mainzer/etal:2012, nugent/etal:2015, nugent/etal:2016, masiero/etal:2017}. Figure 1 of \citet{mainzer/etal:2012} for the 3-band cryo
and the Oct 2010 through Jan 2011 2-band data, Figure 4 of \citet{nugent/etal:2015}, Figure 4 of \citet{nugent/etal:2016}, and Figure 4 of \citet{masiero/etal:2017} for the NEOWISE reactivation 2-band data all present versions of this plot.
The latter three papers give sigma of 14, 20, and 12.5\%
with samples sizes of 90, 53, and 95.  Combining the variances from  these papers gives
an overall scatter  of $\pm 15$\% for the two-band NEOWISE diameters relative to the ROS diameters.
As M2018b notes,
the two band data is the more difficult case for finding diameters, so the 4 band
cryo mission diameters should be better than 15\%.

We note that in the process of compiling the results for Table 1 of \citet{masiero/etal:2011} and Table 1 of \citet{mainzer/etal:2011b}, the full list of calibrator objects was incorporated in an effort to be consistent with \citet{mainzer/etal:2011a}. However, late in the referee process for \citet{mainzer/etal:2011a}, the referee requested that objects with maximum lightcurve amplitudes larger than 0.3 mag be dropped from that paper,
as these are more likely to be highly elongated and thus poor choices for calibrators.  Neither Table 1 in  \citet{mainzer/etal:2011b} nor \citet{masiero/etal:2011} was updated to reflect this reduced calibration set due to an oversight when the calibrator object table was reproduced in these two papers, affecting 0.7\% of objects (3 objects) in \citet{mainzer/etal:2011b} and $<$0.1\% of objects (68 objects) in \citet{masiero/etal:2011}. In order to consolidate all the team's published results into one location, a compilation of them was published in NASA's Planetary Data System for ease of use by the community \citep{mainzer/etal:2016}. A future release of NEOWISE data in PDS will update this and other known errata.

\begin{figure}
\plotone{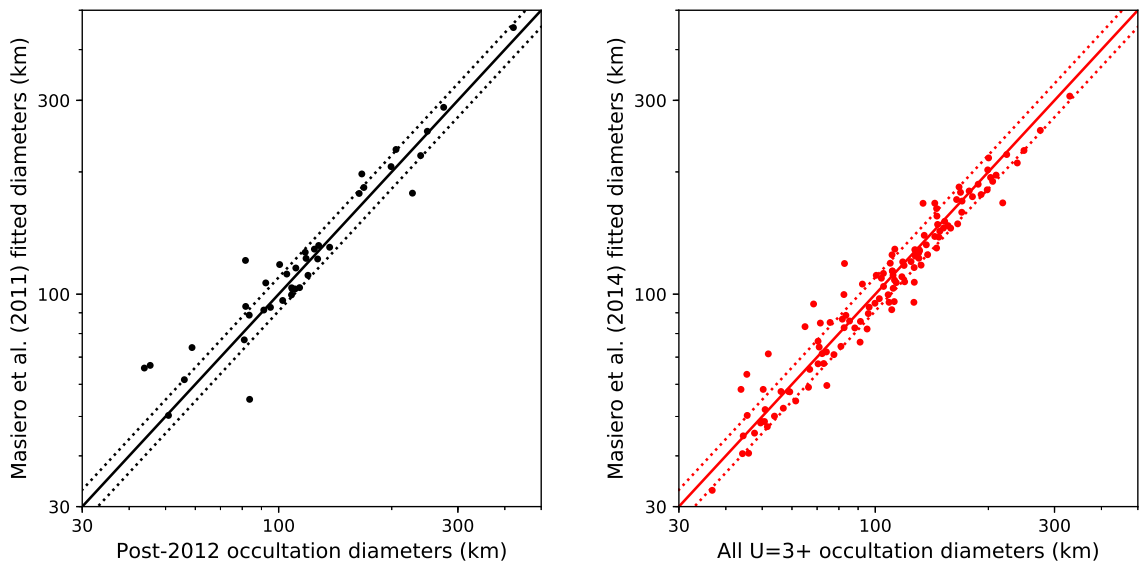}
\caption{Left: Diameters that were calculated solely using data from WISE's 2010 fully cryogenic mission phase in \citet{masiero/etal:2011} are plotted
against new occultation diameters measured in 2012 and later
from \citet{dunham/etal:2017}.  Right: Diameters taken from \citet{masiero/etal:2014} that were computed for calibrator objects using only WISE data from the 2010 fully cryogenic mission phase are plotted against all available occultation diameters. For occultation data, only the highest-quality measurements (U=3 or 4) were used \citep{herald:2018}. Lines show a 1:1 relation and $\pm 10$\% deviations. \label{fig:occDiamComp-1}}
\end{figure}

Since radar, occultation, and spacecraft observations of asteroids continue to be made, it is
possible to make an independent assessment of the WISE diameters using new
occultation data taken since the original calibration paper was written in 2011.  A search of \citet{dunham/etal:2017} found 81 Main Belt asteroids (MBAs) with new
occultation data with quality codes of 3 or 4, indicating the highest-quality data for which robust size estimates are possible\footnote{From PDS \citep{dunham/etal:2017}: \emph{``Quality code for fit:  0 -  not fitted: The quality of the observations is insufficient to allow a reliable fit to either the asteroid's diameter or position;  1 - time only: The observations are sufficiently reliable to permit the determination of the position of the asteroid relative to the star, to a precision of half the asteroid's diameter;  2 - poor: The observations are sufficient to permit the determination of the position of the asteroid to a fraction of its diameter, and to
place some meaningful limits on the size and shape of the asteroid;  3 - good: The observations allow a good determination of the size and orientation of a best-fit ellipse;  4 - excellent: There are many observations around the limb of the asteroid, showing detail in the shape of the asteroid's profile."} Per \citet{herald:2018}, fit codes of U$<$3 are not recommended for use as calibrator objects.}, from occultation events in 2012 or later.  
Of these, 61 also have diameters
determined by WISE during the fully cryogenic mission, and 45 were not used for the original 2011 calibration, giving us the opportunity to verify that the original calibration is still valid in light of these new direct diameter measurements.  These 45 objects are plotted as black dots in Figure \ref{fig:occDiamComp-1}.  Note that diameters determined from multiple epochs using WISE data were averaged; multiple
occultations were averaged; and when an ellipse was fitted to the occultation
data, the diameter is taken to be the diameter of a circle with the same area.  All of the WISE fitted diameters are
from the online version of Table 1 in \citet{masiero/etal:2011}.
The median diameter ratio is $D_\mathrm{WISE}/D_\mathrm{occ} = 1.04$ with the 
16$^{th}$ and 84$^{th}$ percentiles at 0.93 and 1.15.
\citet{masiero/etal:2014} redid the thermal model fits for 2835 MBAs (including nearly all ROS calibrator objects) using a version of NEATM that has the ability to determine albedos in both bands W1 and W2 separately based on the method described in \citet{grav/etal:2012}, and there are no ROS diameters in Table 1 of \citet{masiero/etal:2014}.  The WISE diameters from Table 1 of \citet{masiero/etal:2014} are plotted \vs\ all occultation diameters with quality codes U=3 or 4 as red dots in Figure \ref{fig:occDiamComp-1} for the 58 objects in common. 
The median ratio of the diameters is $D_\mathrm{WISE}/D_\mathrm{occ} = 0.97$ with the 
16$^{th}$ and 84$^{th}$ percentiles at 0.91 and 1.09. Our calibration used only those objects with the highest confidence levels (U=3 or 4), whereas M2018b used all occultation data, including the low confidence levels (U=1 and 2) that are not recommended for use as calibrators, thus likely overestimating the radiometric diameter uncertainty.

Thus, publicly available data allow an assessment of the  \citet{masiero/etal:2011}
and  \citet{masiero/etal:2014} diameters that is independent of the calibration set
used by \citet{mainzer/etal:2011b},  and the observed scatter is consistent with
errors of about $\pm 10$\% $1 \sigma$ for the ensemble of objects.  

\begin{figure}[tbp]
\plotone{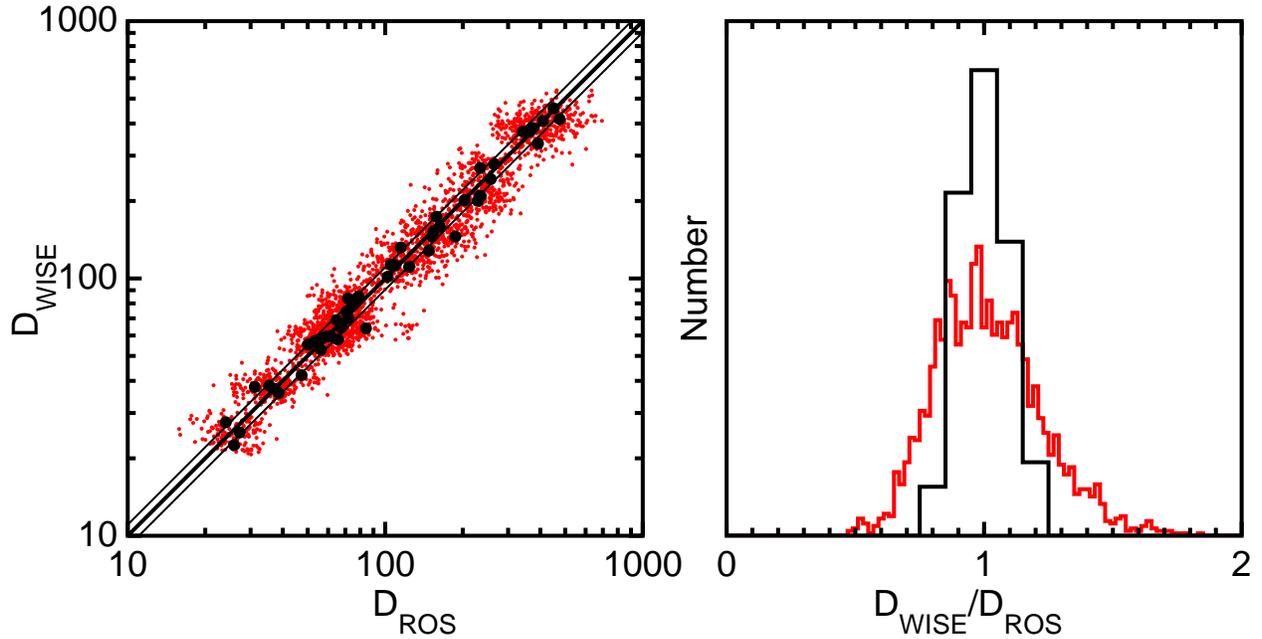}
\caption{Left: A simulation was performed that created ``true" diameters uniform in log between 20 and 500 km for 50 objects; 8\% errors on the true diameters were imposed to create a simulation of the WISE diameters with 8\% systematic diameter errors. Next, four separate values simulating diameters measured by radar, occultation, or spacecraft (ROS) visits were created, each with 12\% errors. These four diameters were averaged to compute a single value for $D_{ROS}$ with 6\% errors.  These average ROS diameters are plotted as black points. Next, the approach of M2018b was simulated by creating 50 Monte Carlo trials per object, where $D_{WISE}$ had random 8\% errors applied, and for $D_{ROS}$, 12\% random errors were applied to a randomly chosen one of the four subvalues. These Monte Carlo values are shown as red dots; each black dot is surrounded by a cloud of red dots. Right: Histograms for the direct comparison of $D_{WISE}$ \vs\ $D_{ROS}$ are shown in black; comparisons based on the non-standard M2018b approach are shown in red. The black histogram produces the expected 16th and 84th percentile ratio of $D_{WISE}$ \vs\ $D_{ROS}$ of 0.88 and 1.08, respectively, whereas the red histogram produces 16th and 84th percentile ratios of 0.82 and 1.21. Thus, the M2018b approach produces an incorrect 20\% scatter.  \label{fig:myhrvold-example}}
\end{figure}

M2018b has Figures 7 \& 8 that at first glance look similar to Figure \ref{fig:occDiamComp-1}, but with much larger central 68\% confidence intervals.  This is caused by two non-standard  procedures used by M2018b.  The first non-standard procedure is to plot not $D_{ROS}/D_{WISE}$ but rather ratios of Monte Carlo samples drawn
from the published means and error estimates.  If the published errors are correct, this
has the effect of doubling the variance (see Figure \ref{fig:myhrvold-example} for an example).  The second non-standard procedure is to combine
multiple determinations of the same value by adding their probability distributions 
instead of multiplying them.  This has the effect of further increasing the errors.
For example, if there are two ROS diameters of $90 \pm 10$ km and $110 \pm 10$ km the
standard procedure gives a diameter of $100 \pm 7$ km, while M2018b 
used a probabilty density function 
$\propto [\exp(-0.5((D-90)/10)^2) + \exp(-0.5((D-110)/10)^2)]$
which has a mean of 100 but a standard deviation of 14 km which is a variance 4 times 
larger than the correct variance.
When combined with the first non-standard procedure this greatly increases the width of the confidence
intervals in Figure 7 \& 8 of M2018b.

Thus, we have shown that the original estimate of a minimum systematic diameter uncertainty for WISE diameters with two or more thermally dominated bands of $\sim$10\% is correct, and the approach of M2018b is incorrect.

\clearpage

\subsection{Accuracy of the WISE/NEOWISE Models}
M2018b asserts that the WISE diameters are not accurate, but that paper only provides thermal fit parameter outputs for two out of the $\sim$150,000 object dataset. Therefore, we cannot fully assess the quality of the M2018b thermal fits; nor does M2018b compare the results to calibrator datasets. Nevertheless, we have established that there are errors in the \citet{myhrvold:2018a} and M2018b model based on examination of the thermal fit results provided for those two objects; the infrared albedos published for both objects are not physically plausible and are inconsistent with the plotted thermal model results that work presents in its Figure 3. Described below is a detailed accounting of the differences between observed and model magnitudes for two asteroids shown in M2018b's Figure 3; they result from a known and fixed issue in the NEOWISE thermal model, as well as an apparent issue or allowance of unphysical parameters on the part of M2018b. 

\begin{figure}
\plotone{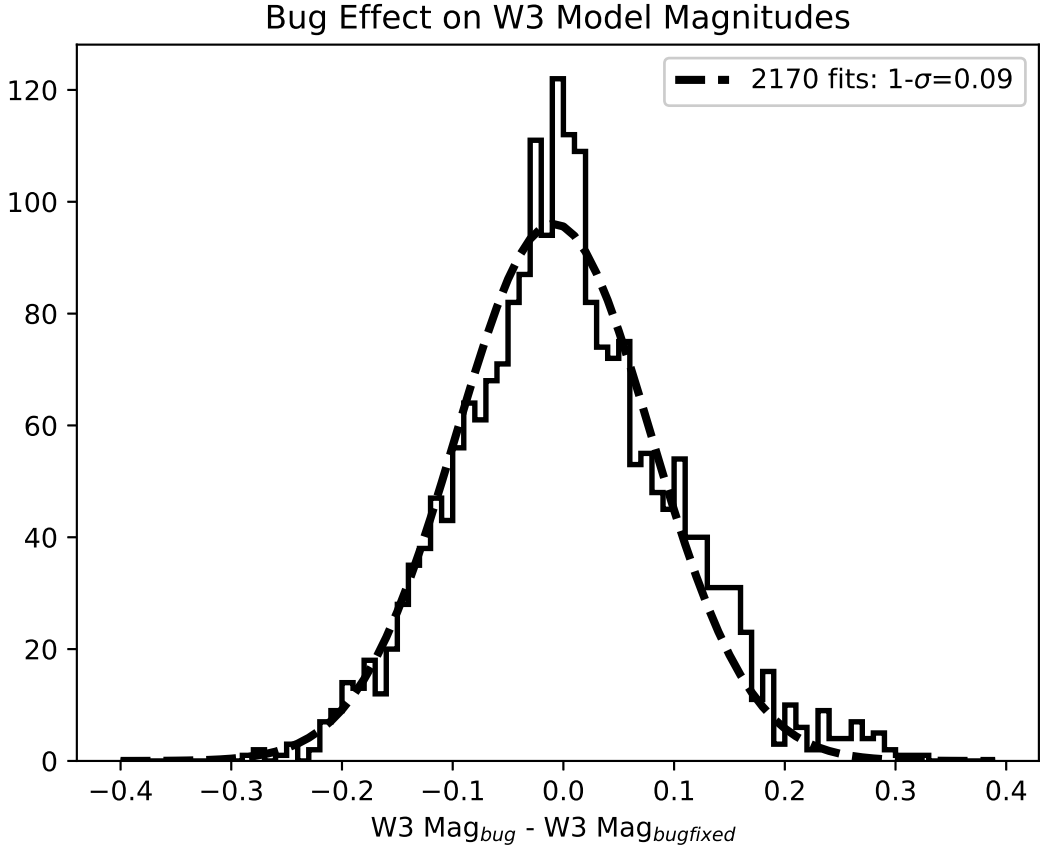}
\caption{Running the NEATM with and without the software issue that flipped a fraction of the facets shows that the majority of objects show offsets in W3 magnitudes that are less than $\sim$10\%. Since diameter error scales as one-half of flux error, the change to the \emph{diameters} is below the quoted minimum systematic diameter error of $\sim$10\% 1-sigma for the ensemble of objects. Although only W3 is shown here because it is the band in which asteroids detected by WISE are brightest, the issue affected all other bands similarly.  \label{fig:W3_offsets}}
\end{figure}

In 2011, a software issue was identified and corrected in the NEOWISE thermal modeling software; the net effect of this issue was to vary the diameters by a few percent on average. Based on analysis of a subset of the data, the magnitude of the shifts was below our quoted minimum systematic uncertainty in diameter that results from using the NEATM (see Figure \ref{fig:W3_offsets}) and was thus determined not to materially change the conclusions of the affected papers.  Because the effect of the issue in general is smaller than e.g. the effects of incomplete coverage of  lightcurve amplitudes, the team was more focused on quantifying the effects of the lightcurve sampling on the derived diameters, and description of the issue was not published after it was remedied; this was an oversight. This issue was present in \citet{mainzer/etal:2011a, mainzer/etal:2011b, masiero/etal:2011, grav/etal:2011}, but was fixed in \citet{grav/etal:2012, mainzer/etal:2012, masiero/etal:2014, mainzer/etal:2014, nugent/etal:2015, nugent/etal:2016, masiero/etal:2017, masiero/etal:2018}. 
 
 The NEOWISE thermal code uses a sphere of triangular facets to compute the temperature and flux distributions. The software issue caused the normal vector of $\sim$30\% of the facets to point inward instead of outward. 
However, the subsequent flux offset depended on the orientation of these facets relative to the observer and the Sun, so most of the time, as shown by a sampling of $\sim$10\% of all objects with W3 data in Figure \ref{fig:W3_offsets}, the flux offset is smaller than 30\% and is typically $\sim$10\%. In cases of objects with offsets caused by the issue that were equal to or greater than the object's lightcurve amplitude, some or all of the model magnitudes can be offset from the measured magnitudes by more than the 1-sigma measurement uncertainty. However, since flux scales with diameter squared, the induced error on diameter will be half the flux error, or $\sim$5\%, which is below the minimum uncertainty of $\sim$10\% 1-sigma for the ensemble of objects that we have observed experimentally (as shown in Figure \ref{fig:occDiamComp-1}).

 M2018b Figure 3 states that for two asteroids (the only two for which fits are shown), the model magnitudes derived from the published NEOWISE thermal fit output parameters are offset by $\sim$0.2 mag in all four wavelengths. As shown in our Figure \ref{fig:W3_offsets}, these two objects are extreme examples of the software issue; it has a smaller effect on most objects. As described below, the change to the respective diameters of both asteroids is 6\% and 8\%, which is lower than the minimum 10\% uncertainty we claim for our data for the ensemble of objects.
\clearpage

\begin{figure}[tbp]
\plottwo{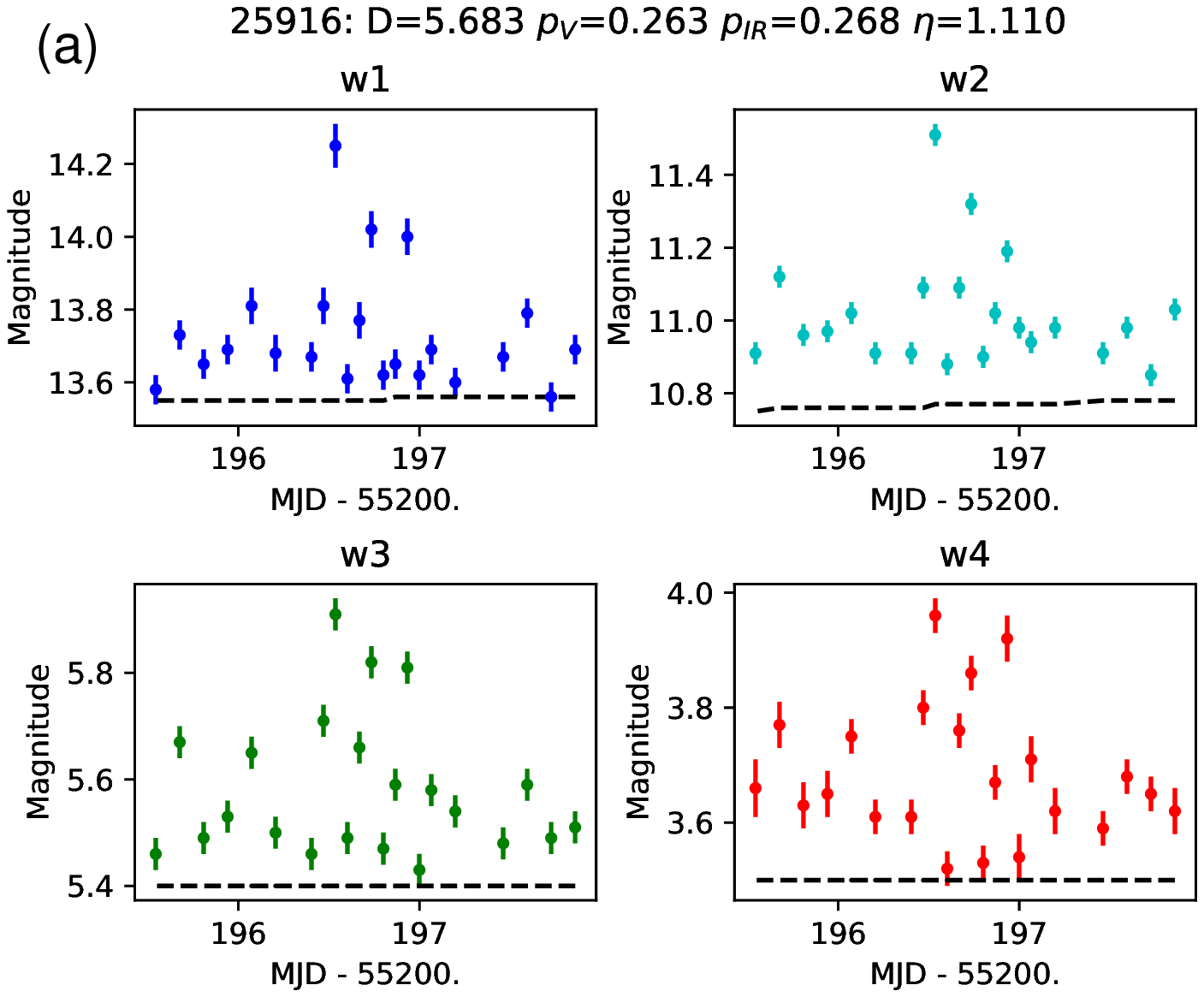}{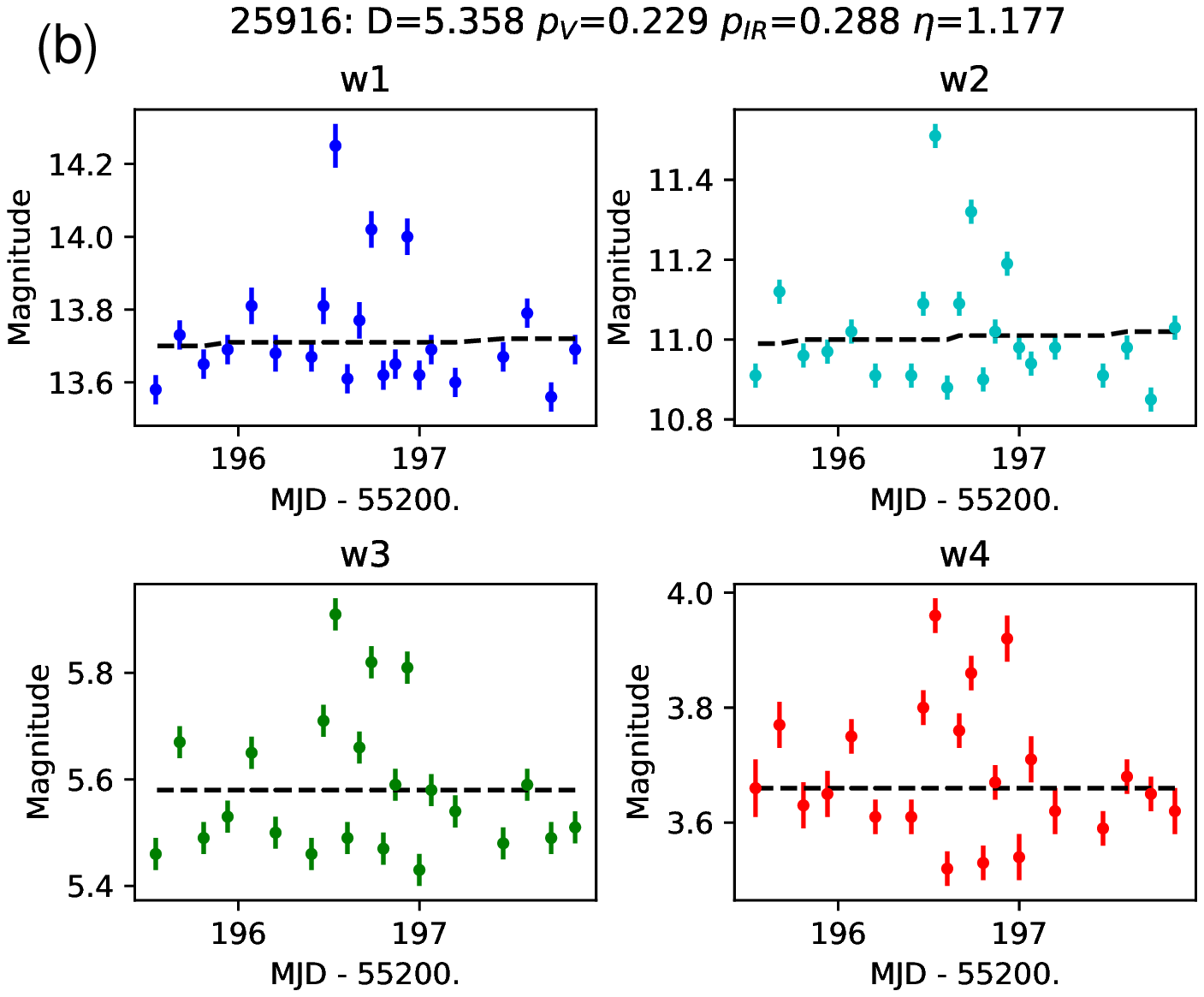}
\plottwo{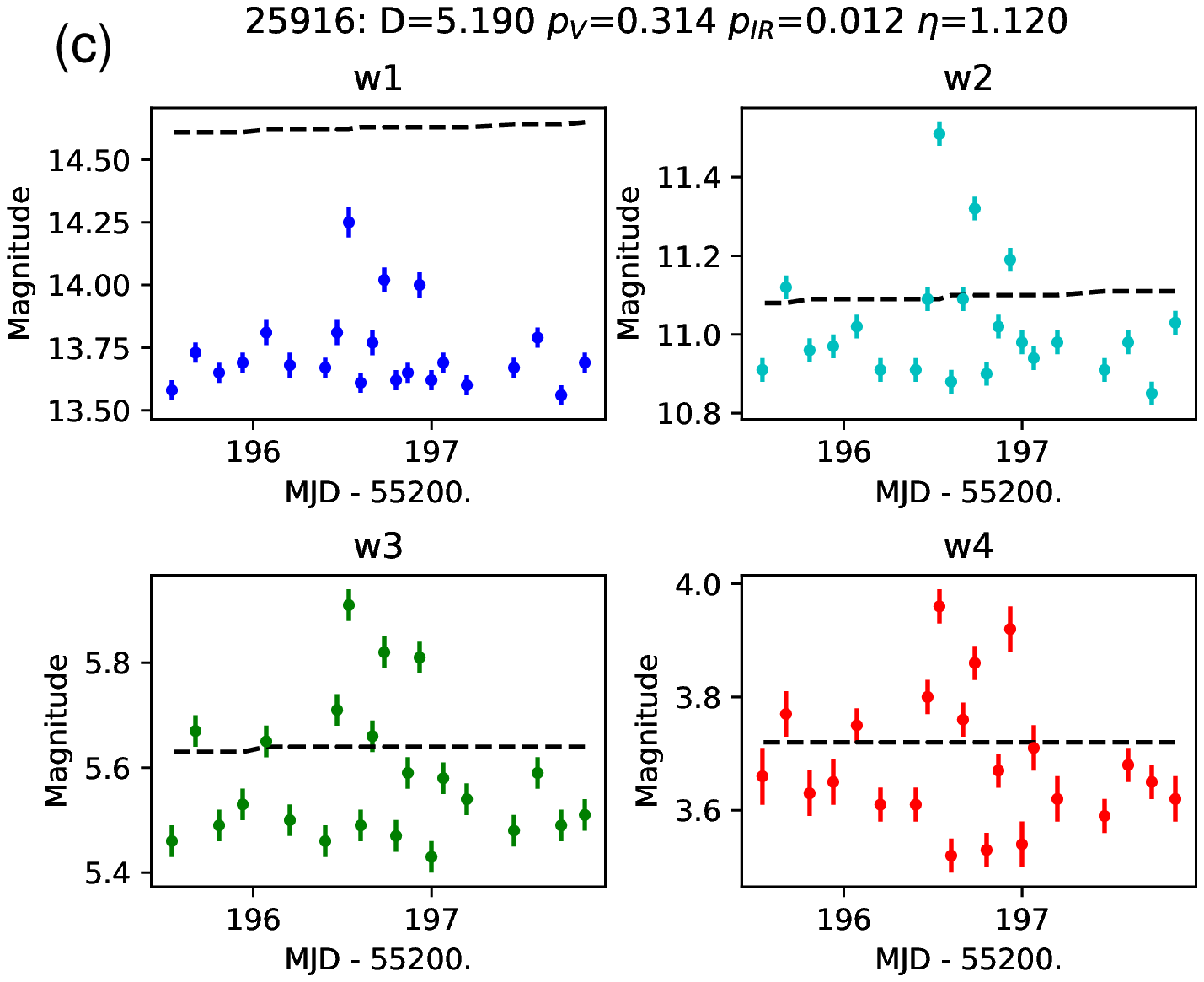}{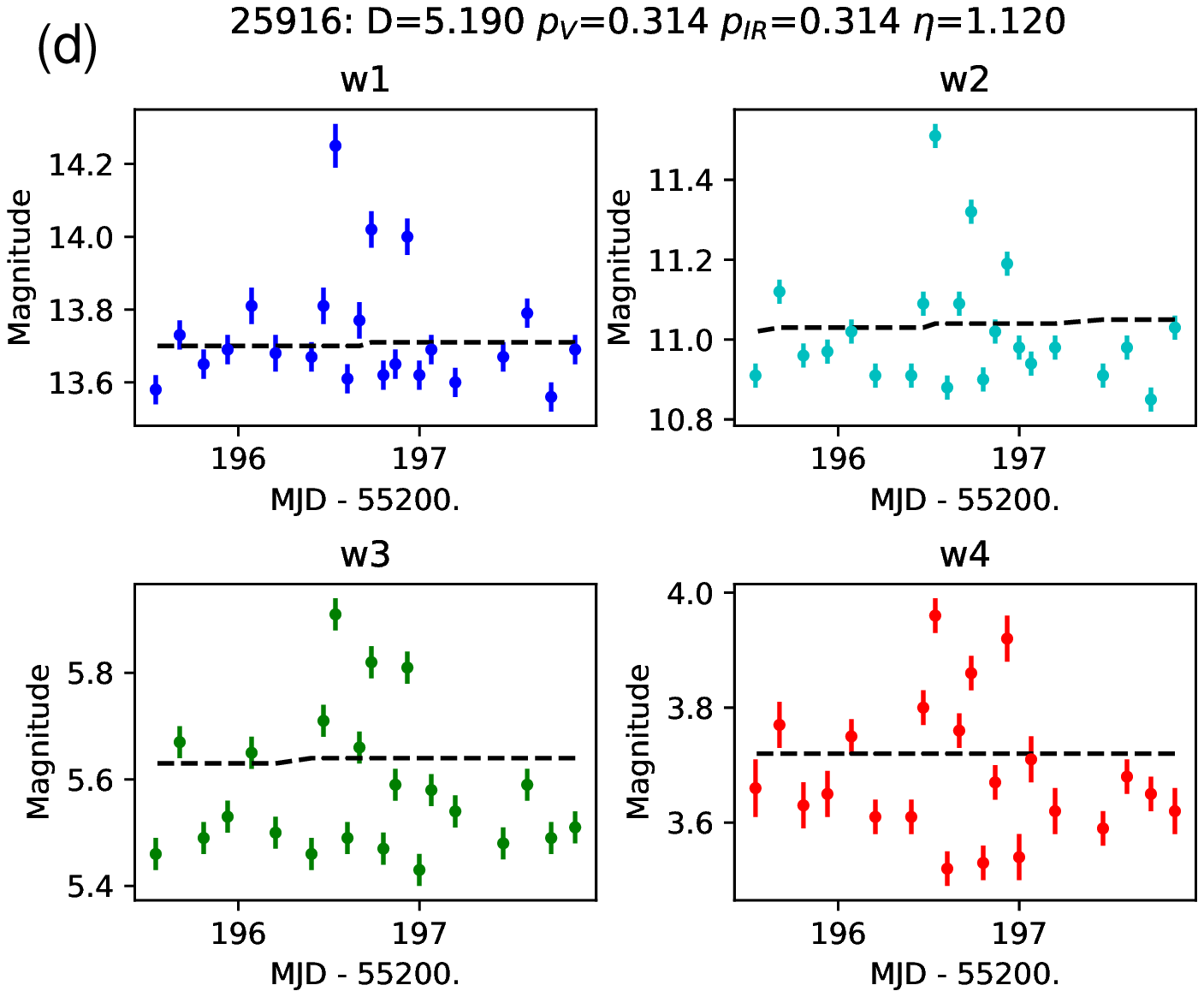}
\caption{a) Thermal fit model magnitudes (dashed lines) are computed for asteroid 25916 using the parameters in \citet{mainzer/etal:2011b} and compared to measured WISE magnitudes; $H$ is taken to be 13.3$\pm$0.3 mag, its value in 2011. b) Updated thermal fit model computed without the software issue shows that the diameter is within 10\% of the diameter for the fit computed in 2011 with the software issue. The current $H$ value of 13.6$\pm$0.3 mag is used. c) Here, a fit is computed using the parameters shown in Figure 3 of M2018b. Using these values, the fit to W1 is $\sim$1 magnitude worse than shown in that paper.  d) Model \vs\ measured WISE magnitudes using the same parameters from M2018b, with the exception that $p_{IR}$ is set such that $p_{V}$=$p_{IR}$=0.314 instead of the value for $p_{IR}$ given in that paper of 0.012, producing a fit that more closely resembles the result shown in Figure 3 of M2018b.   \label{fig:25916}}
\end{figure}

\begin{figure}[tbp]
\plottwo{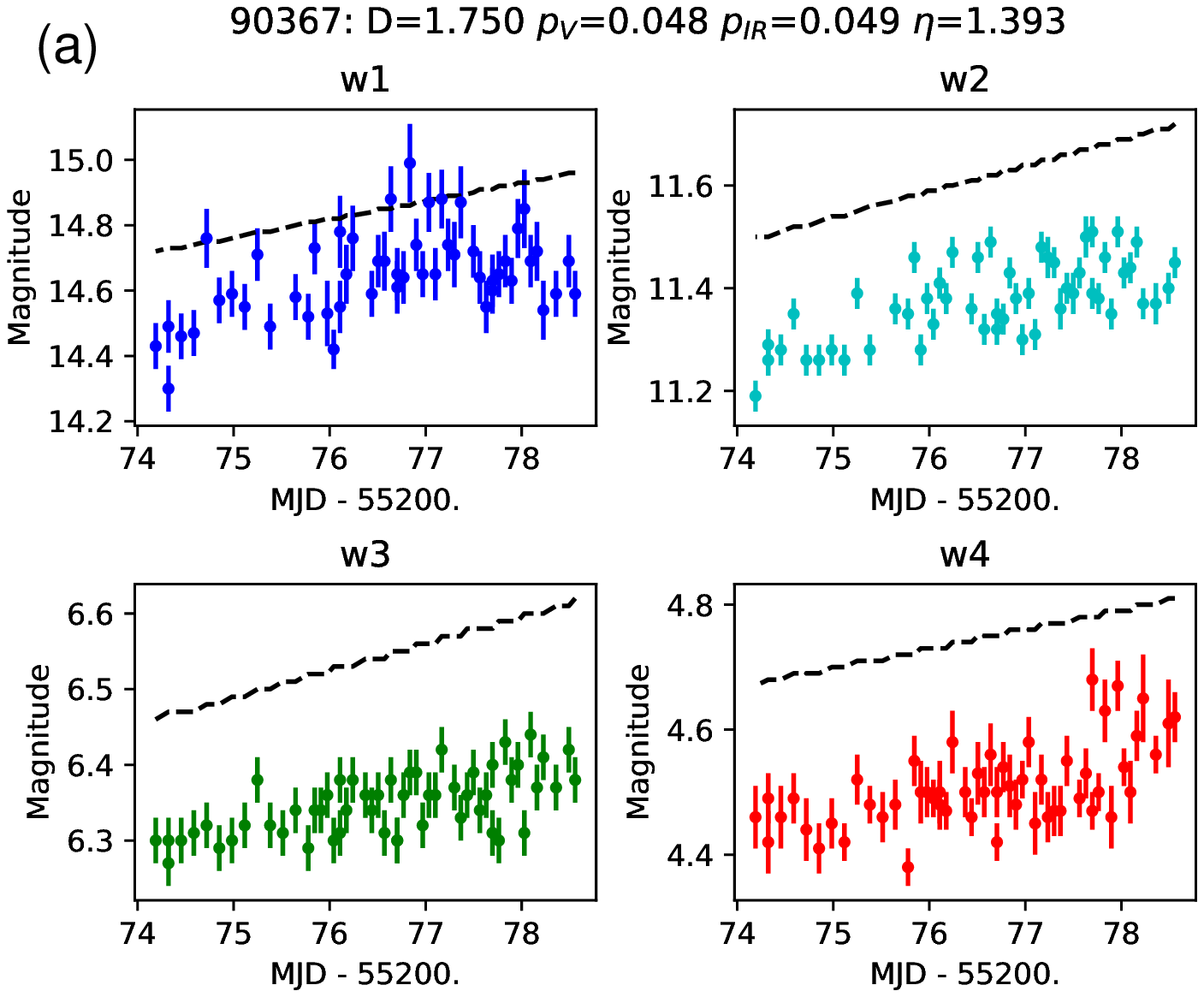}{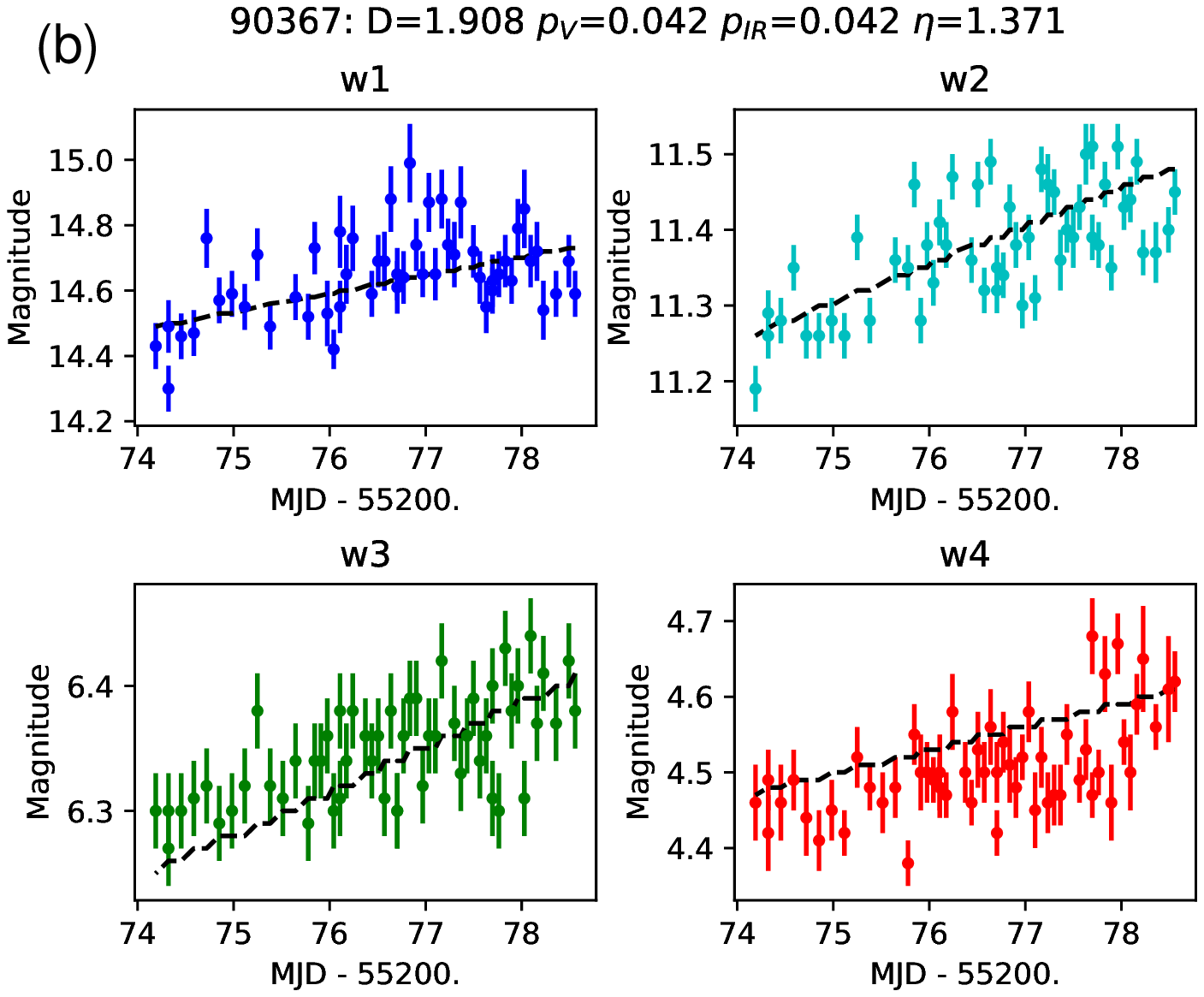}
\plottwo{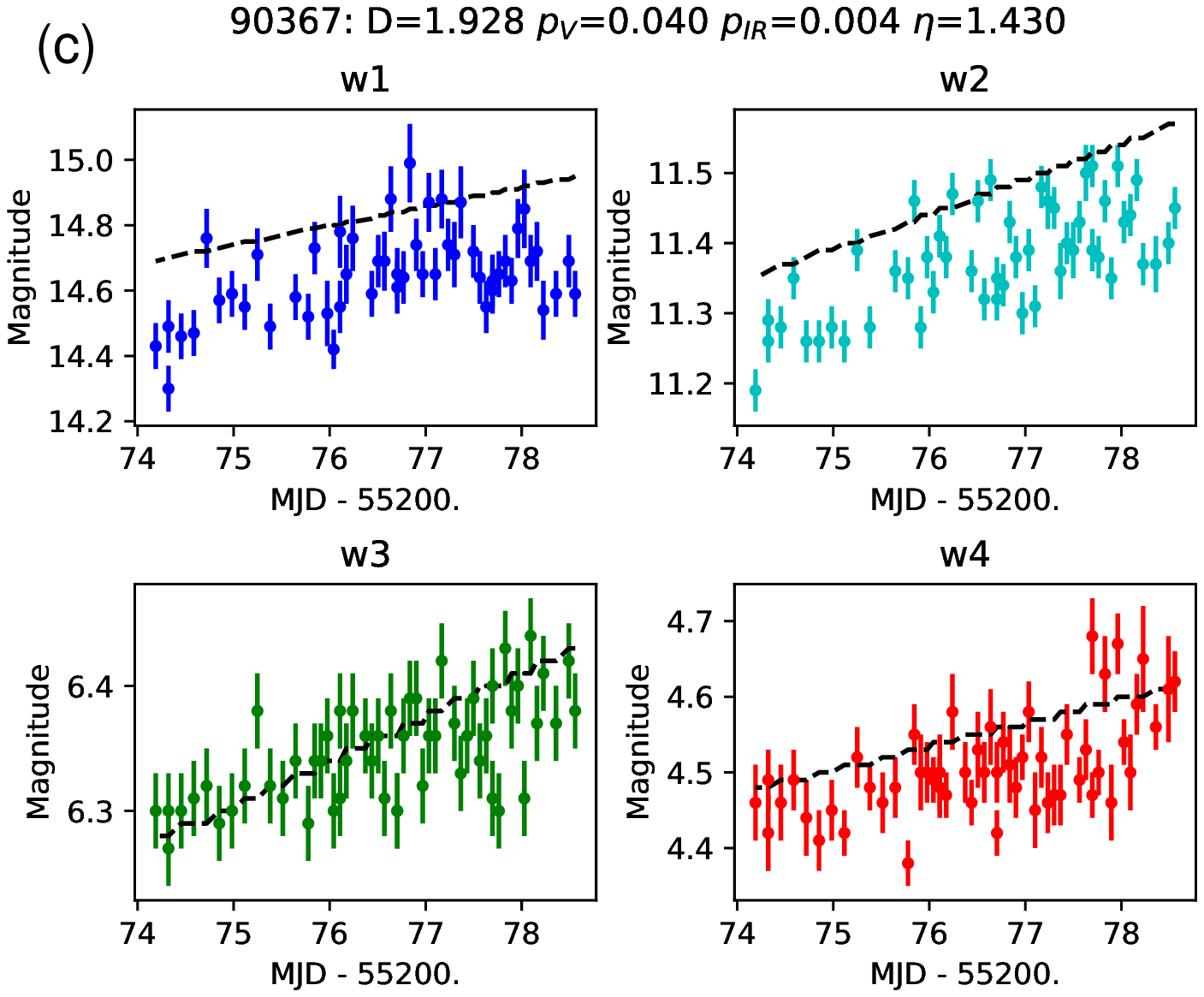}{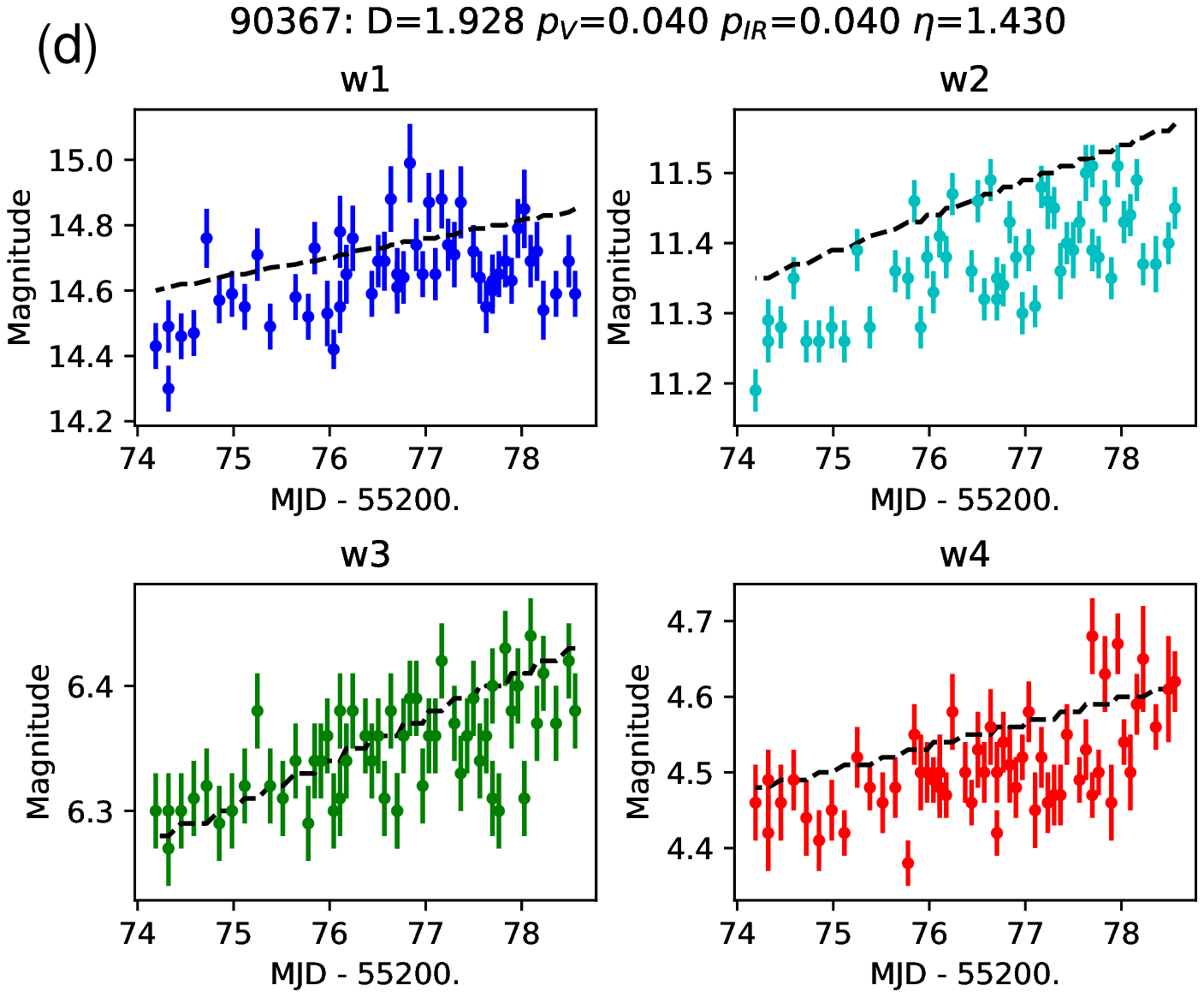}
\caption{a) Thermal fit model magnitudes (dashed lines) are computed for asteroid 90367 using the parameters in \citet{mainzer/etal:2011b} and compared to measured WISE magnitudes. b) Thermal fit model computed without the software issue produces a diameter that is within 10\% of the diameter for the fit computed in 2011 with the software issue. c) Fits using the published values given in Figure 3 of M2018b are shown. An offset between the model and measured W1 magnitudes is evident. d) Fits using the M2018b parameters, but with $p_{IR}$=$p_{V}$=0.040. The offset between model and measurement in W1 is slightly reduced. \label{fig:90367}}
\end{figure}

Figures \ref{fig:25916} and \ref{fig:90367} show fits with (panel a, top left) and without (panel b, top right) the software issue for asteroids 25916 and 90367. Even though the flux offsets for these particular objects with the software issue are $\sim$0.2 mag, the diameters differ by less than the minimum systematic uncertainty of 10\% for both objects (5.68 \vs\ 5.36 km and 1.75 \vs\ 1.91 km respectively) because flux goes as diameter squared. These figures were computed using orbits from the JPL Horizons system for the epochs of the observations. We note that when using the published fit values in Figure 3 of M2018b for 25916, the model flux for band W1 does not go through the measurements (Figure \ref{fig:25916}c). As shown in Figure \ref{fig:25916}d, one possibility for resolving the discrepancy is to assume that the albedo in W1 centered at 3.4 $\mu$m (denoted $p_{IR}$) is instead close to or equal to the visible albedo. Similarly, the offset between model and measured W1 magnitudes for 90367 shown in Figure \ref{fig:90367}c decreases when we set $p_{V}$=$p_{IR}$=0.04 instead of $p_{IR}$=0.004 as shown in M2018b. We cannot replicate the plot shown in M2018b Figure 3 unless we assume a different value for $p_{IR}$ than the published value of 0.004, which is not physically plausible. Spectroscopic measurements indicate that asteroids do not have 3-4 $\mu$m albedos that are a factor of 10-26 lower than the visible albedo \citep[e.g.][]{takir:2012, rivkin:2018, desanctis:2012}, as is reported in Figure 3 of M2018b.   

Also note that the lower panel of Figure 3 in M2018b 
on asteroid 90367 includes two fully cryogenic epochs and two reactivation epochs taken years later,
while the WISE model plotted here is based only on the second fully cryogenic epoch.

\begin{figure}
\plotone{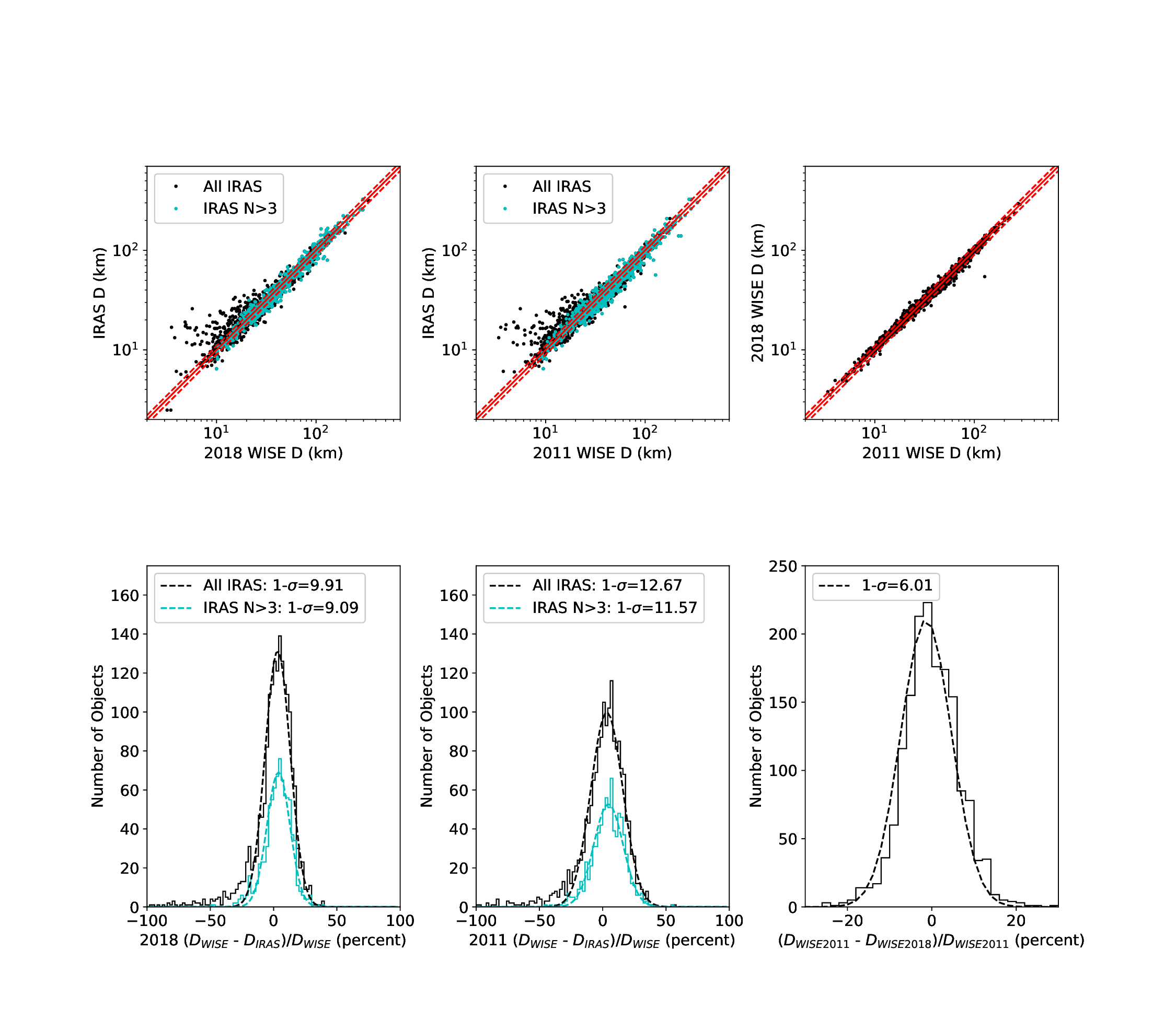}
\caption{a) Diameters from thermal fits run in 2018 on $\sim$1700 objects detected in common between WISE and IRAS are plotted \vs\ IRAS diameters \citep{tedesco/etal:2002}. These fits do not have the software issue, and were run using osculating elements with epochs near the time of observation and magnitudes taken from the WISE All-Sky Single-Exposure Database \citep{cutri/etal:2012}. Red dashed lines indicate 10\% deviations from 1:1 (solid red line) in all plots. Objects with four or more IRAS detections are plotted in cyan in all panels. b) Histogram of the differences between WISE and IRAS. c) Diameters from the 2011 MBA paper \citet{masiero/etal:2011} with all calibration objects removed are plotted \vs\ IRAS diameters, similar to the work shown in \citet{mainzer/etal:2011c}. d) Histogram of the differences between these fits and IRAS. e) Comparison of WISE from panel a (without software issue) and from panel c (with software issue). f) Histogram of the differences between panel a and c WISE fits shows that the flux offsets resulted in diameter errors below the minimum systematic error of 10\% for the ensemble of objects. \label{fig:iras}}
\end{figure}

The resulting diameters are minimally sensitive to  flux offsets due to the software issue, as shown in Figure \ref{fig:iras} (bottom row). The middle two panels show the comparison of fits from \citet{masiero/etal:2011}, which included the flux offsets, \vs\ IRAS. The two datasets agree to within $\sim\pm 12$\% 1-sigma for the ensemble of objects based on the width of the peak in the histogram of diameter differences. Figure \ref{fig:iras} (top row) shows the same dataset now with the corrected model fluxes, yielding a very slightly improved match ($\sim\pm10$\%). When comparing fits done with and without the flux offsets due to the software issue, the difference in derived diameters is lower than the $\sim$10\% 1-sigma minimum systematic error for the ensemble of objects described in \citet{mainzer/etal:2011a, mainzer/etal:2011c}.

\citet{mainzer/etal:2011c} did compare diameters computed using only WISE data to diameters from IRAS and the Midcourse Space Experiment \citep{tedesco_msx:2002} for 1742 objects in common, finding agreement to $\sim\pm 10$\% $1\sigma$ between WISE and previous infrared radiometric diameter surveys, based on the width of the peak in the histogram of diameter differences, although there is a heavy tail in the
histogram for the smallest objects. Figure \ref{fig:iras} reproduces that plot from \citet{mainzer/etal:2011c}, with the objects observed four or more times by IRAS indicated in cyan. Objects with only a few IRAS observations tend to have larger diameters than those derived from WISE data, which typically have 10-12 observations for most objects. When a non-spherical object is detected only a handful of times, it is much more likely to be detected when it is near its maximum projected size, an example of the \citet{eddington:1913} bias.  

Another study on the quantitive assessment of the WISE accuracy was done by \citet{usui/etal:2014}.
 M2018b ignored the main message of \citet{usui/etal:2014}, which
compared IRAS, Akari, and WISE over the 1993 objects in common, and found that
\emph{``the diameters and albedos measured by the three surveyors for 1,993 commonly detected asteroids are in good agreement, and within $\pm 10$\% in diameter and $\pm 22$\% in albedo at 
1$\sigma$ deviation level.''}

\citet{usui/etal:2014} used a straightforward method: they computed a diameter deviation
for each object and each survey: $f_W = d/\overline{d}$, where $\overline{d} = (d_I+d_A+d_W)/3$,
with the $W$ for WISE, the I for IRAS, and the A for Akari.  
They then give in their Table 2 that $\overline{f_I} = 0.992 \pm 0.094$, 
$\overline{f_A} = 1.001 \pm 0.076$, and $\overline{f_W} = 1.006 \pm 0.093$.
One should note that these sigmas slightly underestimate the true variances due to a
correlation between the numerator and denominator.  If we write 
$d_W = (1+\delta_W)d_\mathrm{true}$ and similarly for Akari and IRAS, one finds
three equations for the variances of the three experiments:
\bea
(4/9) \mathrm{var}(\delta_I) + (1/9) \mathrm{var}(\delta_A) + (1/9) \mathrm{var}(\delta_W) & = & 0.094^2 \nonumber \\
(1/9) \mathrm{var}(\delta_I) + (4/9) \mathrm{var}(\delta_A) + (1/9) \mathrm{var}(\delta_W) & = & 0.076^2 \nonumber \\
(1/9) \mathrm{var}(\delta_I) + (1/9) \mathrm{var}(\delta_A) + (4/9) \mathrm{var}(\delta_W) & = & 0.093^2 
\eea
The solution of these equations gives $\sigma_A=0.076$,
$\sigma_I = 0.122$, and $\sigma_W=0.120$.
Overall, \citet{usui/etal:2014}  shows an impressive agreement between three sets of
infrared radiometric diameters from different experiments taking data many years apart,
using different filters, and different analysis methods.

Another example is \citet{ali-lagoa/etal:2013}, which redid NEATM analyses for over 100
B-type asteroids.  They found agreement to within $\pm 5$\% with the NEOWISE diameters from 2011 (which include the software issue).
Since the same data were used, this agreement only tests the consistency
of the computer codes and assumptions about the parameters.  \citet[][Figure 2]{hanus/etal:2018}  also shows excellent agreement between the NEOWISE team's NEATM-derived diameters and diameters computed using an independent thermophysical model that makes use of shape information.

Finally, the tightness of the albedo distribution within asteroid families can be used to limit
the uncertainty of IR radiometric diameters.  \citet{masiero/etal:2018}
show that the intrafamily scatter in albedo in asteroid families is 35\%.  This includes
the uncertainty in the H magnitudes (28\% for $\sigma(H) = 0.3$ mag), any true
variability of albedo within families, and twice the uncertainty of the diameters.
Hence the diameter error is certainly less than 15\%.

Thus we have shown that in spite of a minor software issue that affected early publications, the published WISE diameters are good to within the quoted minimum systematic uncertainty in effective spherical diameter of $\sim$10\%.

\section{Dispelling Misconceptions About the WISE Data and Thermal Modeling}

\subsection{WISE Photometric Measurement Uncertainties}

\begin{figure}[tb]
\plotone{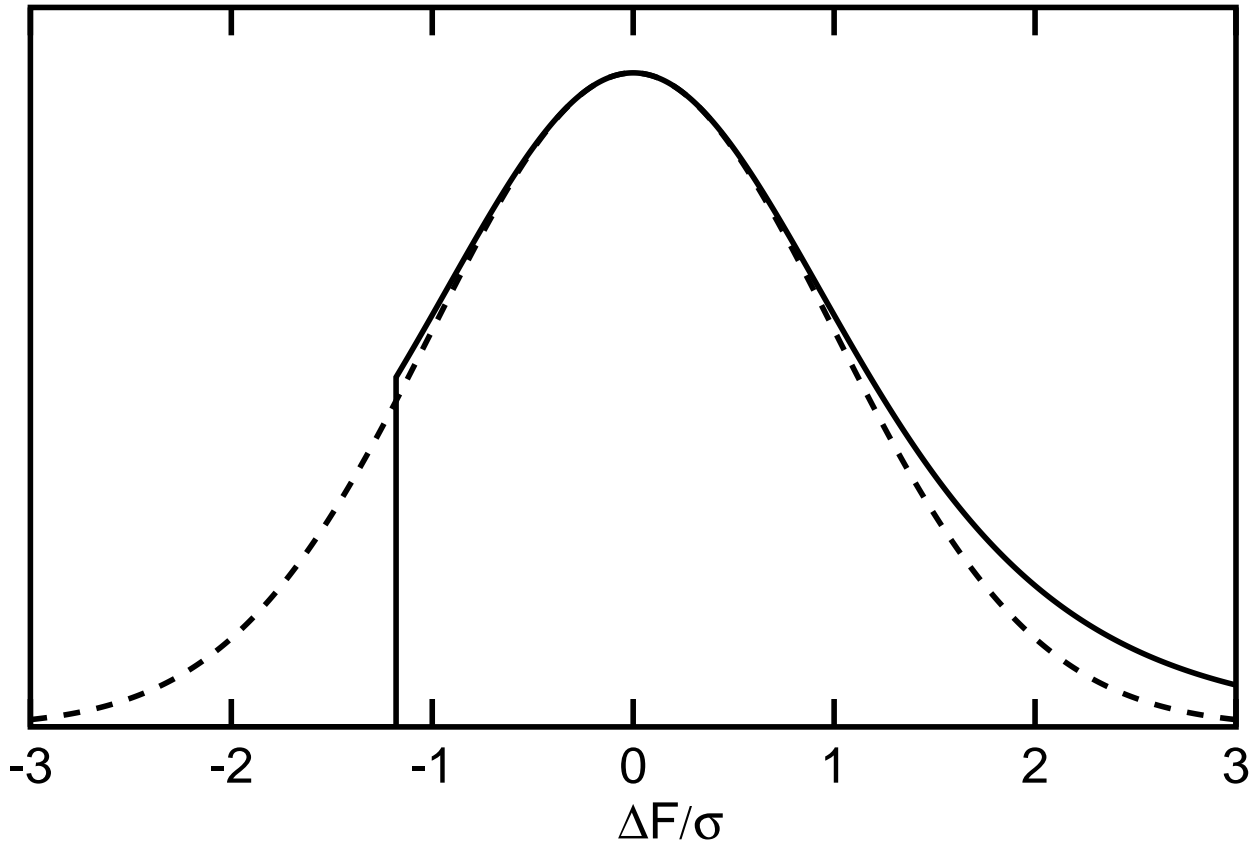}
\caption{A Gaussian [dashed curve] compared to a more realistic distribution of flux errors
with heavy tails due to charged particle hits, transient hot pixels, and other glitches.
The source is assumed to have a true flux 1.18$\sigma$ higher than the flux limit for
the database, so large negative noise excursions never make it into the database.
\label{fig:censored-pdf}}
\end{figure}

M2018b devotes considerable effort to assess whether or not the
WISE brightness measurements have Gaussian errors, asserting that the measurement uncertainties reported by the WISE pipeline are underestimated.  However, there is no reason to expect the errors to be Gaussian because of many effects that cause outliers in real astronomical data. This is true of all astronomical data, regardless of the survey or data source.  First among these is the effect that comes from using the source detection database, which is truncated on the
low flux side.  
To be included in the database, a detection has to have a signal-to-noise ratio (SNR) $> 4$ in a multi-wavelength
detection step, denoted MDET.  These detections are then passed to the photometry module, denoted WPHOT, for accurate photometry
and astrometry  \citep{cutri/etal:2012,cutri/etal:2015}.
For sources with actual fluxes close to the detection threshold, measurements on the negative side of the error dispersion will cause a failure of the MDET threshold test, and the observation will not be included.
This truncation will occur differently in W3, which has the highest SNR
for asteroids of the four bands that are measured and included in the database even when
the in-band SNR is very low or negative.  The magnitudes reported in the database switch to
an upper-limit mode when the SNR is less than 2, but the $wNflux$ (N=1..4, indicating the WISE bandpasses) and 
$wNsigflux$ values (in DN) still give the actual SNR. Figure \ref{fig:censored-pdf} shows a theoretical distribution of flux errors with tails that are larger than Gaussian due to various glitches such as charged particle hits, hot pixels, etc., and a truncation of the low flux side of the distribution due to the detection limit. This flux deviation distribution is obviously not Gaussian, and thus the flux difference distribution is not expected to be Gaussian.

Second-order effects include confusion with other sources
which typically follow a flux distribution that is close to a power law.
Transient hot pixels, charged particle hits, and slight image trailing due to spacecraft
settling also produce heavy tails that make the
flux error distribution non-Gaussian.

By looking at the differences between pairs of asteroid observations taken 11 sec apart, 
which occur in the frame-to-frame overlap region, \citet{hanus/etal:2015} and M2018b
attempted to get clean estimates of the uncertainty in the single frame fluxes reported by WISE.
A typical main belt asteroid will move by 0.1 arcsec in 11 seconds (4 to 5 years to go 360$^\circ$).
This motion amounts to 3\% of
a WISE pixel and will occasionally change the decision as to whether or not to deblend a source,
which leads to a much greater change in the reported flux. M2018b did not specify whether these blended sources were filtered from the
asteroid detection pair data to avoid these problems. These effects also contribute to the non-Gaussian flux distribution.

Now that we have demonstrated that the photometric measurement uncertainties should \emph{not} be Gaussian, we investigate the claims in M2018b and \citet{hanus/etal:2015} that the quoted uncertainties are too small. The flux \vs\ flux plots in Figure 1 of  M2018b generally show good agreement
between asteroids fluxes measured 11 seconds apart.  But both 
\citet{hanus/etal:2015} and M2018b claim that the differences are larger than expected
from the sigma values reported in the single frame detection database.

To get a very dense sample of data to investigate these claims, we have collected data on stars close to the
ecliptic poles that were observed dozens of times during the 4-band cryogenic mission. Stars are appropriate to use for this analysis because, like nearly all asteroids detected by NEOWISE, they are point sources being measured by the pipeline using the same point spread function fitting routines. We require  the following criteria to be satisfied using parameters that are all available in the Single-exposure Source Database: frame quality $qual\_frame$ must be at least 5 to avoid trailed images;
the blending flag $nb = 1$ to avoid problems with source blending; 
$ccflags$ = `0' for the band being tested to avoid contamination by artifacts;
the time from the last detector anneal $dtanneal$ should be $> 2000$ sec for the W3 and W4 bands; WISE must
not be in the South Atlantic Anomaly ($saa\_sep$ $> 0$); and $rchisq$ $< 10$ to throw out
most particles hits and resolved sources.
We also require that the background measurement annulus is not severely truncated
by the edge of the frame,
so sources within 50\asec\ of the frame edge were not used.
For each star we then compute both the rms scatter among the
individual frame fluxes, $wNflux$ (N=1..4), and the average of the $wNsigflux$ values.  
The $wNsigflux$ values come from a model for the detector noise that includes dark current,
sky background, Poisson noise from source photons, and the actual spatial and flux distributions of pixels 
that went into each source extraction.

The dense sample of stars allows us to examine the actual measurement scatter over a wide range in source flux. First, we compute the rms of the flux distribution for each star, and then compute the mean of the rms values in each flux bin as a measure of the characteristic dispersion in flux measurements. Second, we compute the mean of the  $wNsigflux$ values reported by the noise model in the WISE database. These two quantities are compared to confirm the accuracy of the noise model. Thus, we can evaluate whether the noise model reproduces the observed scatter at all flux levels,
or whether it fails for some fluxes.

\begin{figure}[tbp]
\plotone{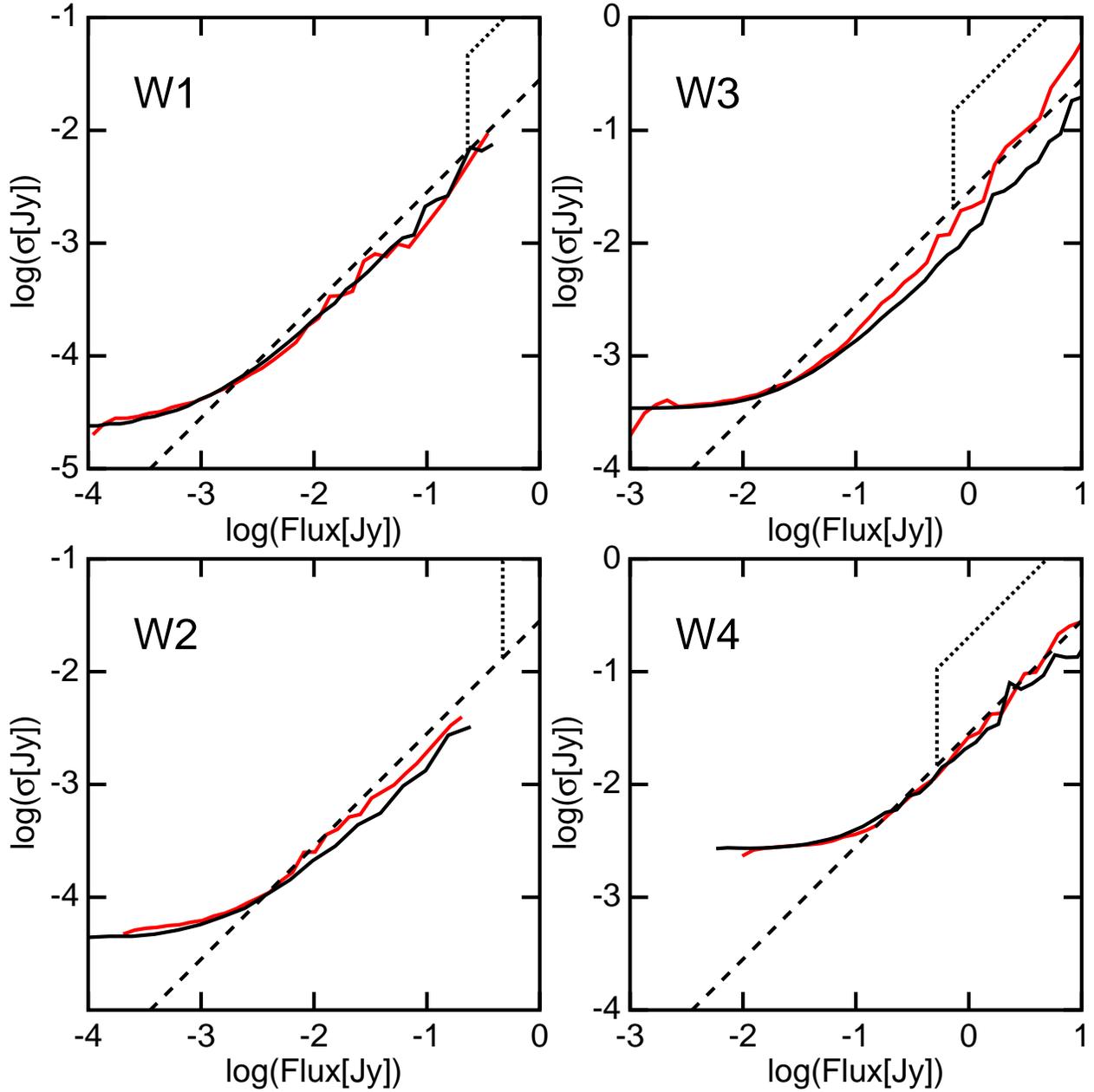}
\caption{A comparison of the observed rms scatter of individual frame flux values for un-blended
stars [red curves] with the average flux uncertainty reported by the noise model in the individual frame detection
database [black curves].  The black dashed line shows the 0.03 mag minimum uncertainty
used in NEOWISE fits, while the black dotted line shows 0.2 mag uncertainty.  In actual use, the  0.2 mag uncertainty was only applied to W3. These plots demonstrate that disagreements between fluxes of unblended stars and those reported in the individual frame detection database are always smaller than the minimum magnitude uncertainties assumed in the NEOWISE thermal model (see text).\label{fig:scatter-sigma-flux}}
\end{figure}

The results of this study are shown in Figure \ref{fig:scatter-sigma-flux}. The noise model is in good agreement with the observed scatter except for
bright and saturated sources in W3.  To account for the observed repeatability of very bright sources, \citet{mainzer/etal:2011b} set any magnitude measurement uncertainty less than 0.03 mag equal to 0.03 mag for all four WISE bands following \citet{wright/etal:2010} and Section VI.3.b Table 1 of the WISE Explanatory Supplement \citep{cutri/etal:2012}. 
To account for the effects of the onset of saturation, \citet{mainzer/etal:2011b} set the measurement uncertainty for saturated sources to 0.2 mag for W3 only.  Saturation was assumed
for W3 $< 4$ mag and W4 $< 0$ mag. 
W1 and W2 are not saturated for asteroids due to their very red colors, but Figure \ref{fig:scatter-sigma-flux} shows
where saturation occurs.  Figure \ref{fig:scatter-sigma-flux} shows that while the flux uncertainties reported in the data are 
underestimated for saturated sources in W3, the actual scatter is always less than the uncertainty
used in the NEOWISE thermal model fits. Our results with a much
larger sample of repeated observations of stars show that underestimated uncertainties
only occur at high signal levels on saturated sources. Thus, the photometric measurement uncertainties provided by the WISE pipeline are accurate, provided that users of the data adhere to the guidelines for handling very bright and/or saturated asteroidal sources given by \citet{mainzer/etal:2011b}.

The NEOWISE team used a Monte Carlo approach to propagate errors from
the input data to the final thermal model results.  Each individual measurement was used in the Monte Carlo analysis that was performed using a least squares fit algorithm for each object. This is an approximate method to
evaluate the linearized propagation of error equation for
a statistic $D$ that is a function of independent random inputs $x_i$:
$\Var(D) = \Sigma_i (\partial D/\partial x_i)^2 \Var(x_i)$, where $\Var$ is the variance.
Note that this equation applies to variances as long as the
$x_i$ have variances, whether or not the distributions are Gaussian.
Thus using a Gaussian Monte Carlo to evaluate the propagation
of statistical errors does not require that the inputs
have Gaussian distributions; using the Gaussian Monte Carlo model of errors is a reasonable approach to constrain the uncertainty on the fit.

\begin{figure}[tbp]
\plottwo{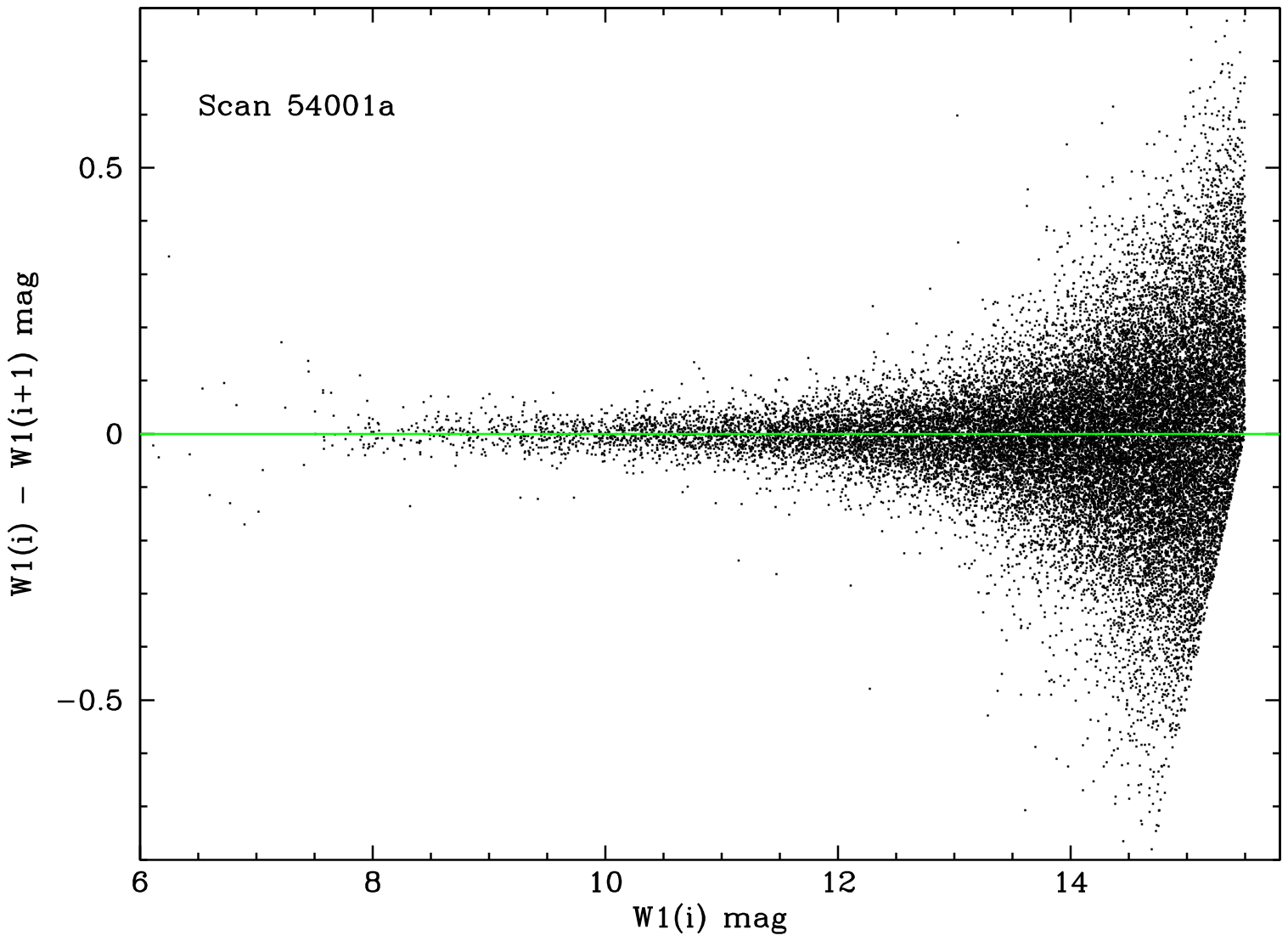}{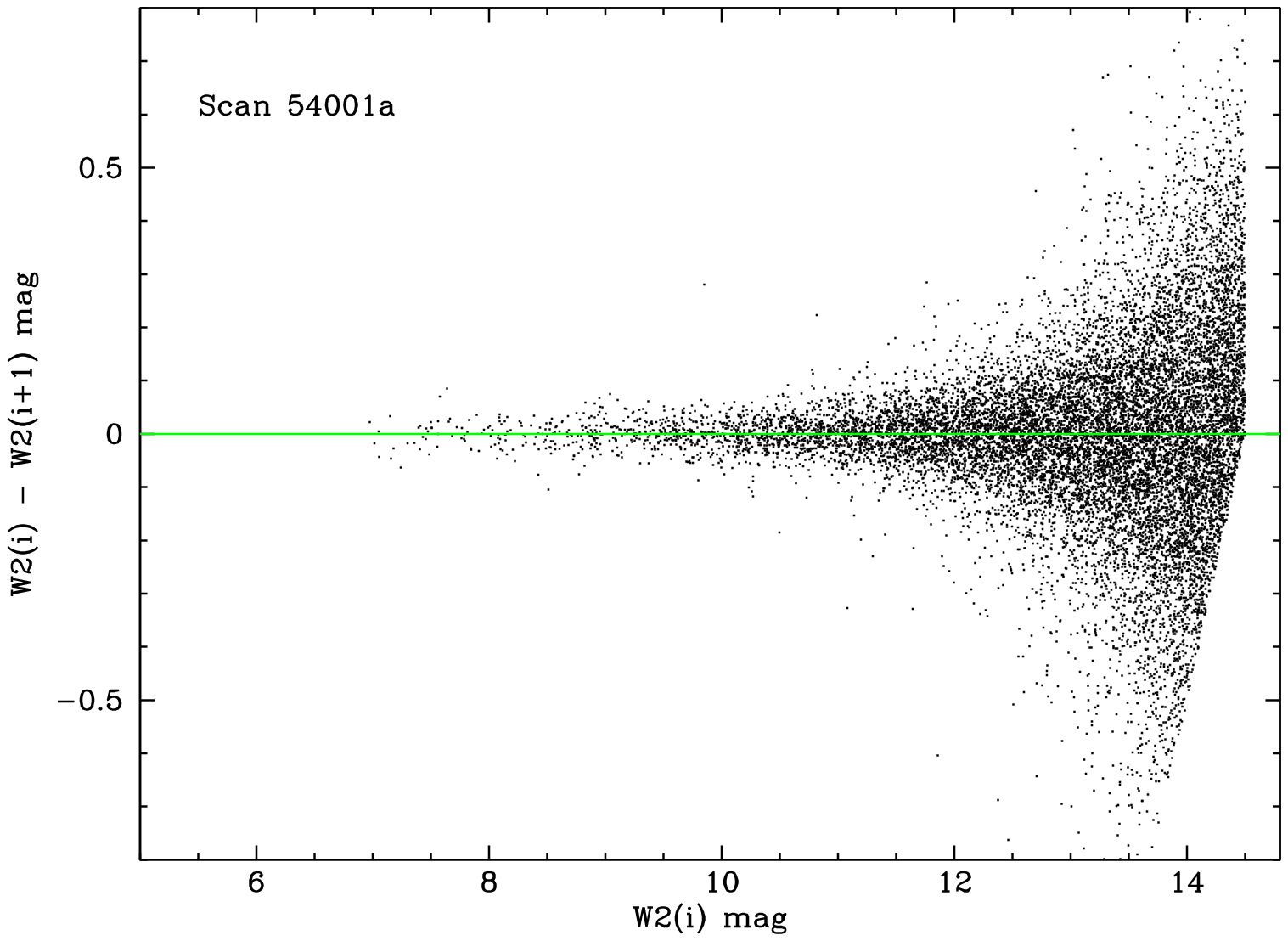}
\caption{A comparison of magnitudes of objects appearing in frame overlap regions illustrates that the majority of these successive observations do not have identical fluxes.\label{fig:delta_mags}}
\end{figure}

Finally, M2018b states ``The WISE pipeline appears to have changed for the
[reactivation] mission; the majority of successive overlapping observations have identical 
fluxes ($f_1 = f_2$)''.  This statement is incorrect.
A cursory check of 12 asteroid observation pairs from the
reactivation shows no cases where $f_1 = f_2$ in W2. 
A systematic search of duplicated measurements of stars during  a randomly selected
scan from late 2014, 54001a, gave a sample
of 26,093 detection pairs in W1 and 14,219 pairs in W2 with a tiny fraction
of anomalous pairs having identical $m_1 = m_2$ as shown in Figure \ref{fig:delta_mags}; this tiny fraction is consistent with the expectations from Poissonian statistics.

Thus, we have refuted the claim of M2018b that the WISE measurement uncertainties are improperly estimated and used, and that the WISE pipeline was changed for the NEOWISE reactivation mission.

\subsection{The linear correction for non-linearity}

The All-Sky release Explanatory Supplement, Figure 8 in \S VI.3.c,\\
\url{"http://wise2.ipac.caltech.edu/docs/release/allsky/expsup/sec6_3c.html#cal_bias"},
shows that saturated sources can have a bias because the flux measurement
in the wings of the saturated point spread function is imperfect.
Examination of the behavior for W2 in Figure 8 in the Supplement shows a very clear effect in the trend of stellar color with
brightness; however, asteroids are faint in W2.  A similar trend in stellar color observed for W3 is important for
the majority of asteroids.  A correction to the W3 magnitudes is applied for W3 $< 4$ mag defined
by a straight line in magnitudes, or a power law in fluxes, with
\be
\label{eq:linear}
\mathrm{W3}_\mathrm{corr} = 
\cases { 0.86 \mathrm{W3}_\mathrm{cat} +0.49 & if $\mathrm{W3}_\mathrm{cat} < 4$ mag\cr
\mathrm{W3}_\mathrm{cat} & otherwise}
\ee 
where $\mathrm{W3}_\mathrm{cat}$ is the magnitude from the single exposure detection database
and $\mathrm{W3}_\mathrm{corr}$ is the corrected value that should be used in fitting.
 
In practice this correction for saturated W3 magnitudes has only a small effect
on the computed asteroid diameters, because the increased $\sigma$ of 0.2 mag deweights
W3 relative to W2 and W4, which usually have high SNR when W3 is saturated.

\subsection{Should $H$ be fit exactly?}\label{sec:H}

M2018b asserts that the equation relating $H$, $p_{V}$, and $D$ \citep{fowler:1992},
\be
H = 5 \log \left(\frac{1329\;\mathrm{km}}{D\sqrt{p_V}}\right)\label{eq:H}
\ee
is not satisfied exactly by the \WISE\ model diameter and albedos, and states that it should always be exactly satisfied.
One should note that this is {\em not} the definition of $H$.  $H$ is the visual magnitude
of the object observed at zero phase angle from a distance of 1 AU, when it is
1 AU from the Sun \citep{bowell:1989}.  Thus $H$ is a statement about the {\em measured} visual 
brightness of the object. Because of this definition, $H$ is almost always an extrapolation of measurements taken at non-zero phase, requiring assumptions about the photometric behavior, which is a source of uncertainty for the value.

M2018b contains a fundamental misconception about the role of this equation. Thermal models fit parameters such as diameter, albedo, and beaming to measurements such as the WISE fluxes and optical magnitudes using techniques such as least squares fitting. $H$ is a measurement that, like the WISE fluxes, is used by the least squares fit to constrain the parameters. Diameter and albedo are parameters that are varied in the model to find the best fit to the measured values ($H$ and the WISE fluxes). Just as not all flux values are expected to be perfectly reproduced in the best-fit model that has more constraints from measurements than parameters being fit, neither will the $H$ measurement be perfectly fit - a natural consequence of measurement uncertainty. 
The thermal fit model gives the predicted value
for the absolute magnitude, $H_p$, and the term $((H-H_p)/\sigma_H)^2$
contributes to the overall $\chi^2$ when fitting a model.
If there are more data points than parameters being fit, then the minimum $\chi^2$
model will have residual errors which will be distributed among all the
observations, with the largest errors typically occurring for the least
well-measured observations: this is usually $H$ as the errors
on $H$ are typically $\sigma_H = 0.2$ to $0.3$ mag.

\begin{figure}[tbp]
\plotone{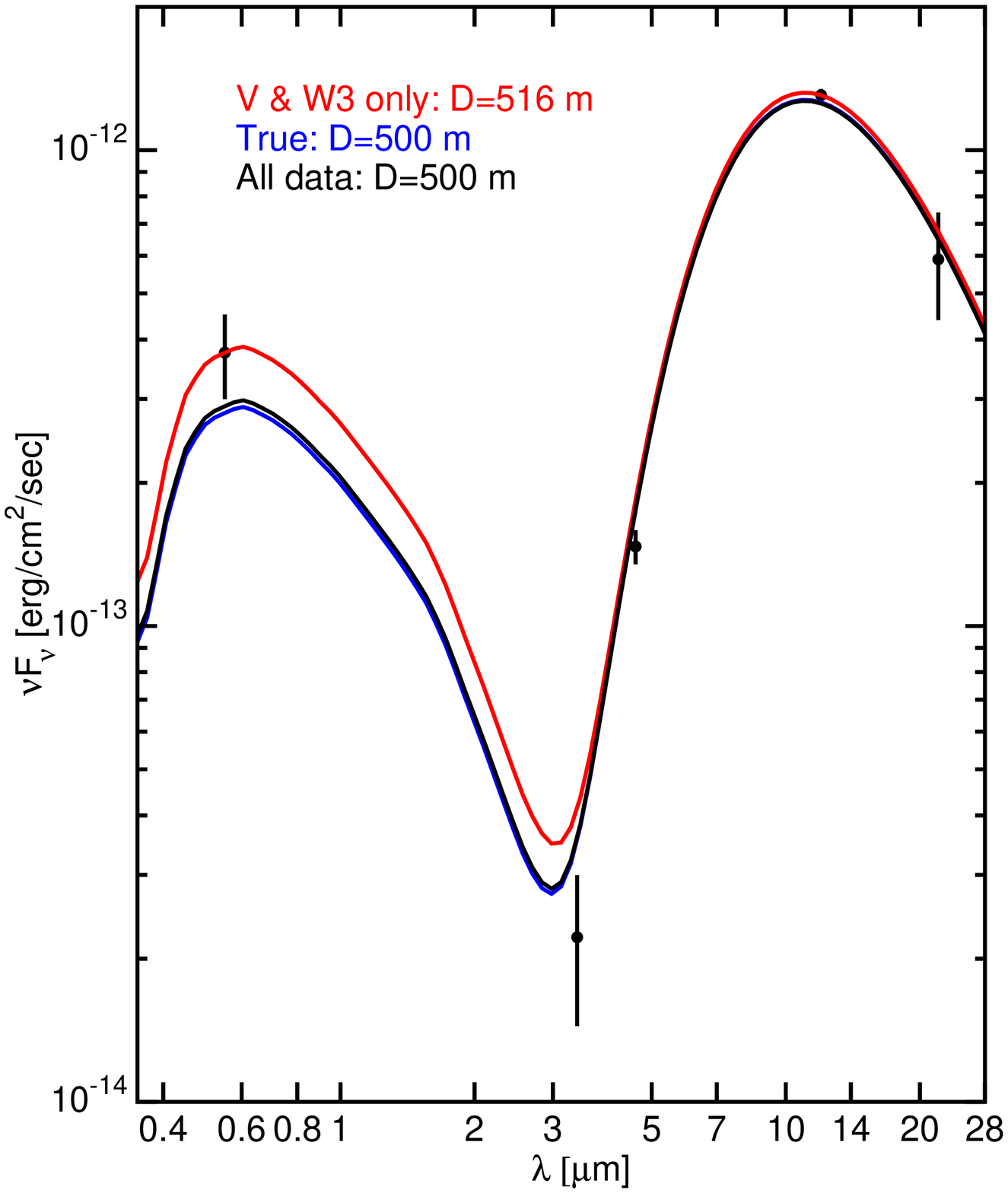}
\caption{A simple model of an isothermal disk with a Lambertian scattering surface.
The blue curve is the true spectrum which shows the non-blackbody
nature of the Sun including the H$^-$ opacity minimum. The black data points
have had Gaussian flux errors applied.  The black curve is a two parameter
fit to all 5 datapoints.  The red curve is a two parameter fit to only the 0.55
and 12 $\mu$m data.\label{fig:disk-model-spectra}}
\end{figure}

\begin{figure}[tbp]
\plotone{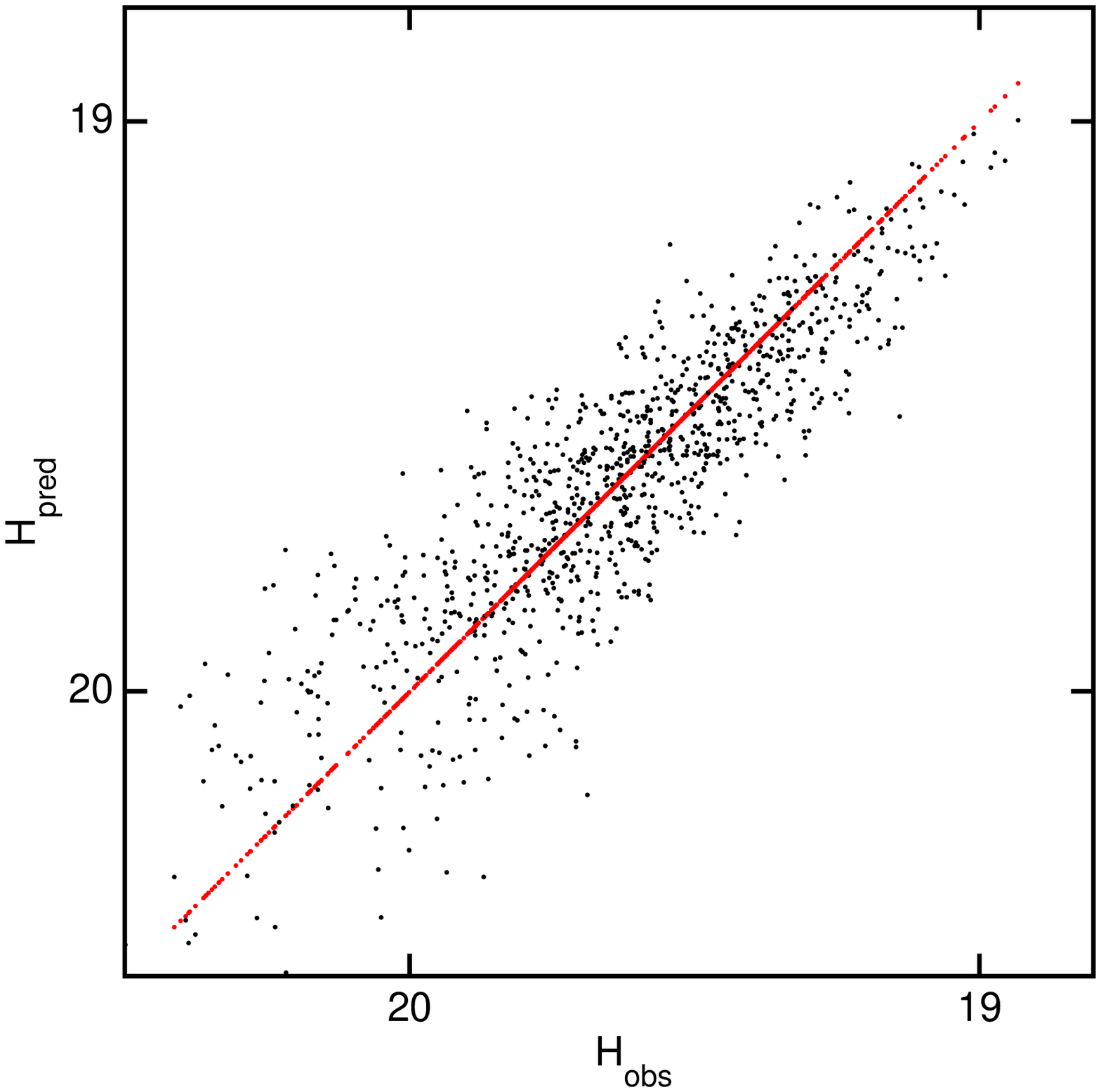}
\caption{A scatter plot from 1000 Monte Carlos showing the 
observed H computed from the V-band flux and the H computed from
the model parameters $p$ and $D$.  The black points are for least
squares fits to all 5 measurement bands (V, W1..4) while the red points use only the 0.55
and 12 $\mu$m data. With only two measurement bands and two parameters, the red points can fit $H$ exactly, but with more measurement bands (V, W1..4), $H$ will not be fit exactly. \label{fig:H-diff}}
\end{figure}

One can see this more clearly by considering a very simple model of a
face-on disk with a geometric albedo $p$ observed at a distance $\Delta$ at 
zero phase angle. The surface is assumed to be a Lambertian scatterer, and to radiate with an angle independent emissivity of $1-p$. The predicted flux at any wavelength is given
by
\be
F_\nu^\mathrm{pred} = 
\left(\frac{\pi D^2/4}{\Delta^2}\right)\left[(1-p)B_\nu(T)+p  
F_\nu^\odot/\pi\right]
\ee
This model is designed to satisfy Kirchhoff's Law and has three
parameters: the albedo, diameter, and temperature.
The temperature is fixed by requiring that the reradiated power
equal the absorbed power.  
Since the disk is Lambertian, the phase integral $q$ is unity.
Since $p$ is assumed to be independent of frequency,
the bolometric Bond albedo $A = pq = p$.
For the V band, the Planck function $B_\nu(T)$ is negligible compared
to the incident flux from the Sun $F_\nu^\odot$ so this equation
reduces to Eq(\ref{eq:H}).  The partial derivatives of the predicted
fluxes with respect to the albedo
$\partial F_\nu^\mathrm{pred}/\partial p$ are all non-zero, so the
minimum $\chi^2$ model in the case where one has five observed frequencies
(visual $H$ magnitude and the four WISE bands, i.e. V and W1..4) will generally not satisfy any of the five observations exactly.
Examples were computed for a disk 1 AU from the Sun and 1 AU from the
observer, with an equilibrium $T = 332$ K for an isothermal two-sided disk.
The blue curve in Figure \ref{fig:disk-model-spectra} shows the true model
spectrum computed with $p = 0.1$ and $D = 500$~m, while the black data points 
have been perturbed by Gaussian
errors with the usual $\sigma(H) = 0.3$ mag and the noise computed from the combination of eight WISE frames (the typical minimum number of frames available on the ecliptic plane in bands W3 and W4)
given by \citet{wright/etal:2010}.  The black curve is the best fit to
all five datapoints.  It does not agree with input value of $H$.
This particular case was chosen as a dramatic example from the
1000 Monte Carlo trials shown in Figure \ref{fig:H-diff}. 

If the number of data points matches the number of free parameters,
then the data points will be fit exactly.  The red curve in Figure \ref{fig:disk-model-spectra} 
shows a fit to only the $V$ band and $W3$ data.  With two parameters
and two datapoints, the fit matches the data exactly.  However, the
errors for $W1$, $W2$ and $W4$ lead to a $\chi^2$ summed
over all five bands that is higher than the $\chi^2$ for the black curve. Repeating this example over 1000 Monte Carlos gives the scatterplot
shown in Figure \ref{fig:H-diff}.  The fits restricted to $H$ and $W3$ only
give the red diagonal line, matching the input $H$ exactly, while the fits
to all five datapoints show a scatter $\sigma_\mathrm{H} = 0.16$ mag, which
is less than the 0.3 mag input uncertainty. 

There is one exception to this rule, which is that if the albedo $p$ has
no effect whatsoever on the predicted infrared fluxes, then Eq(\ref{eq:H})
can be fit exactly since $p$ can be adjusted to fit Eq(\ref{eq:H}) without
any effect on the IR predictions if the diameter is scaled
as $D \propto (1-p)^{-1/2}$.  But the reflected light does affect the
IR fluxes, W1 especially.  In fact \citet{myhrvold:2018a} is titled
``Asteroid thermal modeling in the presence of reflected sunlight'' and makes
the point that the albedo has a significant effect on the predicted IR fluxes,
and one consequence of this is that Eq(\ref{eq:H}) is not satisfied exactly
when one has enough data to do a least-squares fit for the parameters.

In order to better constrain the contribution of reflected light to each band in the NEOWISE NEATM model, the fitting routine was run iteratively, so that the number of constraints on emitted and reflected light was known.  Observations in W3 and W4 are dominated by emission for all observed objects (out to Saturn's orbit); however, the W1 and W2 bands can range from both thermally dominated for NEOs 1 AU from the Sun to both dominated by reflected light for Hilda and Jovian Trojan objects.  Thus an iterative approach is needed, where a notional model is fit and then the reflected light contribution in each band is calculated to determine which of the NEATM model parameters can be constrained.

We have thus refuted the claim of M2018b that Equation \ref{eq:H} should be exactly satisfied.

\subsection{Is the beaming parameter $\eta$ a fixed property of an object?}

While  M2018b argues in its appendix that
$\eta$ is a fixed property of an object, this is untrue.  \citet{wright:2007}
shows $\eta$ \vs\ viewing angles for a single simulated object, and there is a wide
range of $\eta$ from 1 to 4 while changing only the observer location.
\citet{harris/drube:2016} make the  point that changes in sub-solar latitude can change the value of $\eta$.  Therefore, an object can have different values of $\eta$ at different points in its orbit, even if its phase angle and sub-solar temperature are the same.
Moreover, \citet{wolters/etal:2008}, \citet{delbo/etal:2003}, \citet{mainzer/etal:2011b}, \citet{harris/drube:2016}, and \citet{trilling/etal:2016}
all found experimentally that $\eta$ changes with phase angle. 

Thus $\eta$ is a combination of both fixed properties of an object such as thermal inertia
and rotation rate, and observation dependent quantities like the sub-Solar latitude,
sub-observer latitude, and the longitude difference between the Sun and 
observer.  Changing these angles by ten degrees can make a significant difference
in the temperature distribution over the observable face of an object, and hence $\eta$. The only parameters that should be the same at all epochs are those that do not depend on observing geometry, i.e., parameters that are intrinsic properties of the surface. $\eta$ is not an intrinsic property of the surface but rather a model factor that incorporates the effects of several intrinsic and observational properties (this is why we have generally taken the approach of modeling different viewing epochs separately with the NEATM). Thus, we have refuted the claim of M2018b that $\eta$ should be a fixed property of an asteroid.

\section{Conclusion}

We have shown that a number of claims made in M2018b regarding the WISE data and thermal modeling are incorrect. That paper provides thermal fit parameter outputs for two of the $\sim$150,000 object dataset and does not make a direct comparison to calibrator objects. We are unable to reproduce the results for the two objects for which M2018b published its own thermal fit outputs, including diameter, albedo, beaming, and infrared albedo. In particular, the infrared albedos  for the two asteroids published in M2018b are unphysically low; however, using infrared albedos that are similar to the visible albedos, we are able to closely replicate M2018b's Figure 3.

While there are some minor issues with consistency between tables and small offsets  between fluxes published by the WISE/NEOWISE team, as there are in most projects, the team tracks and resolves issues as new tools and methodologies become available in subsequent data releases after suitable review. Moreover, we have shown that these updates do not substantially change the results and conclusions drawn from the data. We have demonstrated (see Section 2.1) that the diameters published to date are accurate to within the minimum  $1\sigma$ uncertainty of $\sim$10\%  that we have quoted for the ensemble of objects detected at low to moderate phase angles in two thermally dominated bands with good SNR and good sampling of rotational lightcurves.   

In addition, we have shown that the WISE measurement uncertainties are accurately reported by the pipeline and can be used verbatim, with the caveat that for saturated objects in the W3 band,
the $w3sigflux$ values are too low, and there is a documented flux bias for saturated objects
that can be
corrected by a linear (in magnitude space) correction for non-linearity.
A similar bias is clearly present in W2 but is not relevant for asteroids.
Readers are advised to use 0.03 mag as the minimum magnitude error, and
not uncritically accept high SNRs for objects with W3 $< 4$ mag, as recommended in \citet{mainzer/etal:2011a}.
The 0.2 mag minimum error used
by the NEOWISE team for sources with W3 $< 4$ mag may be overly conservative, but it is certainly
adequate.

We find that the NEATM is a useful model even though it is obviously a simplified representation of real asteroids, as has been documented in the literature. For example,
the ``beaming parameter'' $\eta$ does not in fact lead to any beaming
when the infrared flux is computed;
for $\eta \neq 1$ the NEATM does not conserve energy; a spherical shape
is assumed; and for high phase angles
the diameters become inaccurate \citep{mommert/jedicke/trilling:2018}.  All of these issues can be addressed 
using a thermophysical shape model \citep[e.g.][]{hanus/etal:2018}, but the
computational load is currently impractically high for the \cal{O}($10^5$) objects
observed by WISE, and most of the objects in the WISE database lack sufficient supporting information such as shape, reliable spin state solutions for thermophysical modeling.  In practice, and as noted by \citet{mommert/jedicke/trilling:2018}, the NEATM gives pretty good results
especially when the phase angle is $< 60^\circ$, so it continues to
be very useful.

Nonetheless, the ``preliminary results'' published to date can be improved. 
With many years of post-cryo operation, many thousands of asteroids have been
observed over multiple epochs, and for many of them there is a high
enough signal to noise ratio to justify multi-epoch analysis using thermophysical models.
The ultimate accuracy of this approach has not been tested as extensively as the NEATM,
but notable successes have been achieved.
One successful case is the 2\% diameter agreement between the pre-encounter 
radiometric analysis of Itokawa based on extensive data and the Hayabusa direct observations 
\citep{mueller/hasegawa/usui:2014}.  Further tests on Bennu and Ryugu will be made
later this year. 

Moreover, the team has always planned a future data release that incorporates all the improvements and lessons learned: the WISE single-exposure data were reprocessed following the publication of the 2011 results using improved calibrations and algorithms developed following the end of the primary mission survey operations and are now available through IRSA; the WISE Moving Object Processing was rerun on the reprocessed data, and additional detections and objects were found and reported to the Minor Planet Center in 2018; better orbits are now available for many objects, along with in many cases better $H$ and $G$ measurements; and finally, new data continue to be collected by the NEOWISE spacecraft as of this writing. All these improvements will be incorporated into updated thermal model fits that will be published in the literature and archived in PDS. 

\section{Acknowledgments}

Part of the research was carried out at the Jet Propulsion Laboratory, California Institute of Technology, under a contract with the National Aeronautics and Space Administration. This publication makes use of data products from NEOWISE, which is a project of
the Jet Propulsion Laboratory/California Institute of Technology, funded by the National
Aeronautics and Space Administration. This publication makes use of data products from
the Wide-field Infrared Survey Explorer, which is a joint project of the University of California, 
Los Angeles, and the Jet Propulsion Laboratory/California Institute of Technology,
funded by the National Aeronautics and Space Administration. We gratefully acknowledge
the services specific to NEOWISE contributed by the International Astronomical Union's
Minor Planet Center, operated by the Harvard-Smithsonian Center for Astrophysics, and
the Central Bureau for Astronomical Telegrams, operated by Harvard University. We also
thank the worldwide community of dedicated amateur and professional astronomers devoted
to minor planet follow-up observations. This research has made use of the NASA/IPAC
Infrared Science Archive, which is operated by the California Institute of Technology, under
contract with the National Aeronautics and Space Administration. We thank our many colleagues who provided helpful commentary to this manuscript.

\end{document}